\documentclass[twocolumn]{aastex631} 
\usepackage{amsmath}

\newcommand\gaia{\textit{Gaia }}

\newcommand\Pcut{$P_{\rm cut}$}

\begin{document}

\title{Different Planetary Eccentricity-Period (PEP) Distributions of Small- and Giant-Planets}
\footnotetext{Corresponding author: Dolev Bashi, db975@cam.ac.uk}

\author[0000-0002-9035-2645]{Dolev Bashi}
\affiliation{Astrophysics Group, Cavendish Laboratory, University of Cambridge, JJ Thomson Avenue, Cambridge CB3 0HE, UK}
\affiliation{School of Physics and Astronomy, Tel Aviv University, Tel Aviv, 6997801, Israel}

\author[0000-0002-3569-3391]{Tsevi Mazeh}
\affiliation{School of Physics and Astronomy, Tel Aviv University, Tel Aviv, 6997801, Israel}

\author[0000-0002-8368-5724]{Simchon Faigler}
\affiliation{School of Physics and Astronomy, Tel Aviv University, Tel Aviv, 6997801, Israel}



\begin{abstract}

We used the database of $1040$ short-period ($1 \leq P < 200$ days) exoplanets radial-velocity (RV) orbits to study the planetary eccentricity-period (PEP) distribution. 
We first divided the sample into low- and high-mass exoplanet sub-samples based on the distribution of the (minimum) planetary masses, which displays a clear two-Gaussian distribution, separated at 
$0.165 M_\mathrm{J}$.
We then selected $216$ orbits, low- and high-mass alike, with eccentricities significantly distinct from circular orbits.
The $131$ giant-planet eccentric orbits display a clear upper envelope, which we model quantitatively, rises monotonically from zero eccentricity 
and reaches an eccentricity of 0.8 at $P\sim 100$ days. 
Conversely, the $85$ low-mass planetary orbits display a flat eccentricity distribution between $0.1$ and $0.5$, with almost no dependence on the orbital period. 
We show that the striking difference between the two PEP distributions is not a result of the detection technique used.
The upper envelope of the high-mass planets, also seen in short-period binary stars,
is a clear signature of tidal circularization, which probably took place inside the planets,
while the small-planet PEP distribution suggests that the circularization was not effective,
probably due to dynamical interactions with neighboring planets.
\end{abstract}


\section{Introduction} 
\label{sec:intro}

The detection of thousands of exoplanets has ushered in a new era in our understanding of planetary systems, revealing a diverse array of planetary architectures \citep[][]{udry23} and compositions that challenge traditional theories of planet formation and orbital dynamics \citep[e.g.,][]{lissauer11,ZhuDong21ARAA,mordasini24,Yang24}. 
A central aspect in the study of these planetary systems is the role of tidal interaction of the planet with the central star and the dynamical interaction with neighbouring planets in shaping the orbits and eccentricities of exoplanets \citep[e.g.,][]{rasio96,MatsumuraRasio10}. 
Therefore, the observed planetary eccentricity distribution attracted much attention in the past, 
as it might help distinguish between the different theories of exoplanet formation and dynamical evolution \citep[see a study with a broader perspective by][]{ke-ting24}. 

For example, \cite{kipping13} and \cite{shabram16} fitted the shape of the eccentricity distribution of planets, either with a $\beta$-function or with a two-component Gaussian mixture model;
see also \cite{Eylen19}, who considered the different eccentricity distributions of single and multiple planets, and \cite{Dong21}, who studied the distribution of planets discovered by TESS in its first year of operation. 

The dependence of the eccentricity on stellar metallicity and multiplicity, and in particular 
on
the planetary mass, was considered by a few studies \citep[e.g.,][]{XieDong16,Mills19,Ribas&Miralda07, he20, An23,Rosenthal24}. 
For example, \cite{kane12} used the transit durations of the Kepler data to conclude that  
the eccentricity distribution is consistent with that found from radial-velocity (RV) surveys to a high degree of confidence, and the mean eccentricity of the Kepler candidates decreases with decreasing planet size, indicating that smaller planets are preferentially found in low-eccentricity orbits. We will come back to this distinction in the present work.

Quite a few works considered the eccentricity dependence on the orbital period (hereafter, Planetary Eccentricity-Period (PEP) distribution). Early works appeared already in 2011
\citep{pont11,pont12}, 
when the number of planetary orbits was quite small. 
Later works \citep[e.g.][]{konigel17, dawson18, Jackson23} have further aimed to confront the observed PEP with the high-eccentricity migration (HEM) scenario, which assumes that giant planets were formed with high orbital eccentricities at large distances from their parent star, and then migrated inward by tidal interaction induced by the high eccentricity \citep{Socrates12, Bonomo17,dawson18, Jackson23}.

A few studies e.g., \cite*{konigel17} and
\citet[][in their review paper, 
Figure 4 in particular]{dawson18}; see also \citep[][]{Jackson23},
focused on the upper envelope of the PEP as an important feature of the planetary population,
which is the subject of this study. 

The number of derived planetary orbits grew substantially in the last few years, especially when high-precision spectrographs dedicated to the RV study of exoplanets went into operation \citep[e.g.,][]{harps-n12,carmenes14,espresso21}. Therefore, it is time to revisit the PEP distribution with the newly accumulated orbits and its dependence on the planetary mass. 
We chose to use only RV orbits as 
the orbital eccentricities and planetary masses are not well constrained by transit light curves alone \citep[but see, for example,][]{kipping08, Ford08, dawson12, kipping14}. We fit an upper-envelope function to the PEP distribution, using a formalism we developed recently \citep*{Bashi23} to analyze the eccentricity distribution and circularization period cut-off of binary stars. Following previous studies, we compare the PEP distributions of the small and giant planets, finding substantially different shapes.

Section~\ref{sec:sample} presents our sample selection aimed to produce an unbiased subset of low- and high-mass planets. Section~\ref{sec:PEP} presents the PEP distribution of the small and giant planets,
section~\ref{sec:discussion} discusses the results and section~\ref{sec:conclusions} summarizes this study.

\section{The sample} 
\label{sec:sample}
For the study of the PEP distribution, we used the Exoplanet Archive 'Composite' catalog \citep{Akeson13},\footnote{https://exoplanetarchive.ipac.caltech.edu/index.html} a comprehensive and well-maintained database that aggregates data from multiple sources, including transit surveys and RV measurements. We opted for this catalog due to its extensive coverage and
reliability, which has been cross-verified against other databases and 
studies
\citep[e.g.,][]{Perryman18, Bashi18, Adibekyan19}. As of the 28th of August, 2023, the catalog includes orbital periods and eccentricity for $1987$ planets that were observed and confirmed by RV measurements.
In this study we focused on the $1040$ 'short-period' $1 \leq P < 200$ days orbits.

While the Exoplanet Archive 'Composite' Planet Data Table offers useful preliminary insights into the population-level properties of exoplanets, such studies require careful scrutiny to ensure statistical robustness. We therefore have conducted a rigorous examination of the sample selection and potential observational biases, as detailed in the following sub-sections. 
This includes 
the difference in exoplanet detection methodologies and the position on the color-magnitude diagram (CMD) of the parent stars. 

\subsection{Division between low- and high-mass planets}

\begin{figure}
	\includegraphics[width=8.5cm]{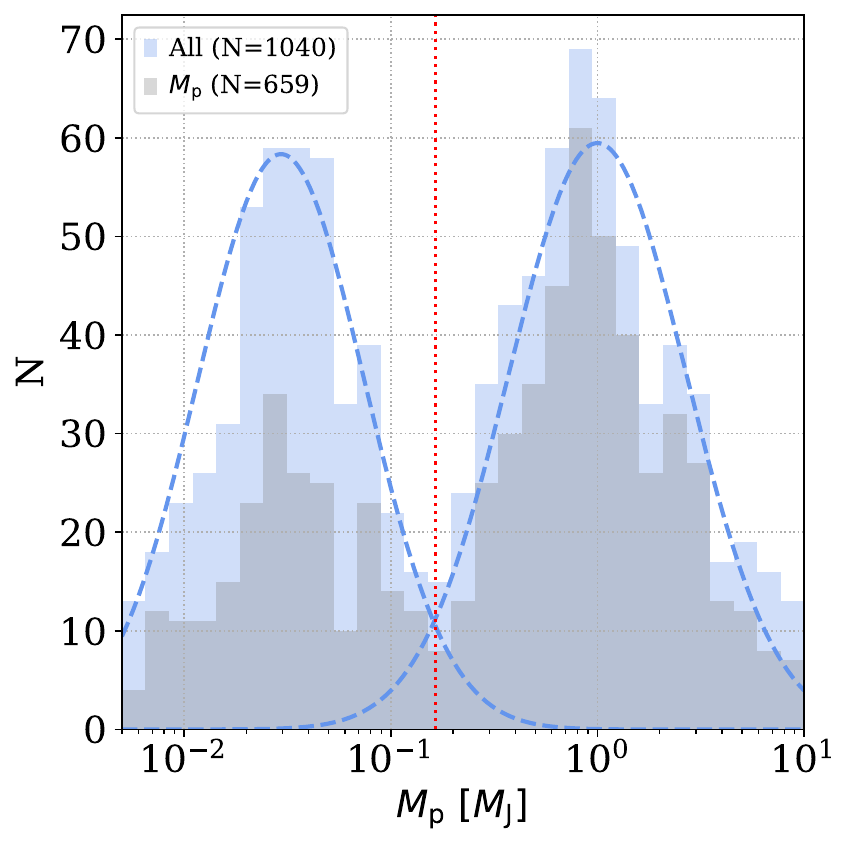}
    \caption{Minimum-mass histogram (blue bins) of all planets with RV orbits, eccentric and circular alike. Dashed curves represent our best Gaussian-Mixture Model of the small- and giant-planet samples. The mean and standard-deviation of the small planets (in log \(M_{\rm J}\)) are $-1.53$  ($=9.4 M_{\rm \oplus}$)
    and $0.40$, and $0.00$ ($= 1 M_{\rm J}$) and $0.43$ for the giant planets. The red dotted line marks the intersecting point between the two Gaussians at $M_{\rm p} = 0.165M_{\rm J}$. Overlaid, using gray bins, is the
distribution of the transiting planets in our sample.}
    \label{fig:PlanetMassHist}
\end{figure}

To divide the sample into small- and large-mass planets, we 
plot in 
Figure~\ref{fig:PlanetMassHist} the (minimum) mass histogram of the planets of all $1040$ RV orbits, eccentric and circular.
The figure clearly reveals two distinct populations of planets, which we have modeled using a Gaussian Mixture Model (GMM) \citep{pedregosa2011scikit}.\footnote{https://scikit-learn.org/stable/modules/generated/\\sklearn.mixture.GaussianMixture.html} One Gaussian component is centered around \(M_{\rm p} \simeq 1 M_{\rm J}\), representing our high-mass (giant) planets. The other Gaussian component is centered at \(M_{\rm p} \simeq 10 M_{\rm \oplus} = 0.03 M_{\rm J}\), representing the low-mass (Super-Earth and Sub-Neptune) planet sample. Here, \(M_{\rm J}\) and \(M_{\rm \oplus}\) denote the masses of Jupiter and Earth, respectively.

A significant result of the GMM fitting is the identification of a division at \(M_{\rm p} = 0.165 M_{\rm J}\), around the Neptunian desert \citep*[e.g.,][]{ida04, mazeh16}, which we use to separate the two populations into $470$ small and $570$ giant planets. 

To show that our division is not dependent on the detection method, we overlay, using gray bins, the distribution of the transiting planets in our sample.
In these cases, the planetary mass estimate derived from the RV orbit, $M_{\mathrm{p}} \sin i$,  is similar to the actual planetary mass, as $\sin i$ is close to unity. 
One can see that the gap and the division between the small and giant planets is similar to the gap derived for the overall sample.

\subsection{Significance of the orbital eccentricity}
\label{sec:ecc}

As pointed out by \cite{LucySweeney71} and more recently by \cite{hara19}
 small orbital eccentricities are not always reliable;
such eccentric orbits might be indistinguishable from circular ones. Consequently, we divided our sample into 'eccentric' orbits, for which
$e > 2 \Delta e$, where $\Delta e$ is the reported lower uncertainty of the eccentricity and 'circular' orbits, for which $e \leq 2\Delta e$, or in cases where the detection paper assumed a circular solution. 
We were left with $762$ circular and $301$ eccentric orbits. 

\subsection{Planet-host stars selection}

\begin{figure*}
\includegraphics[width=18cm]{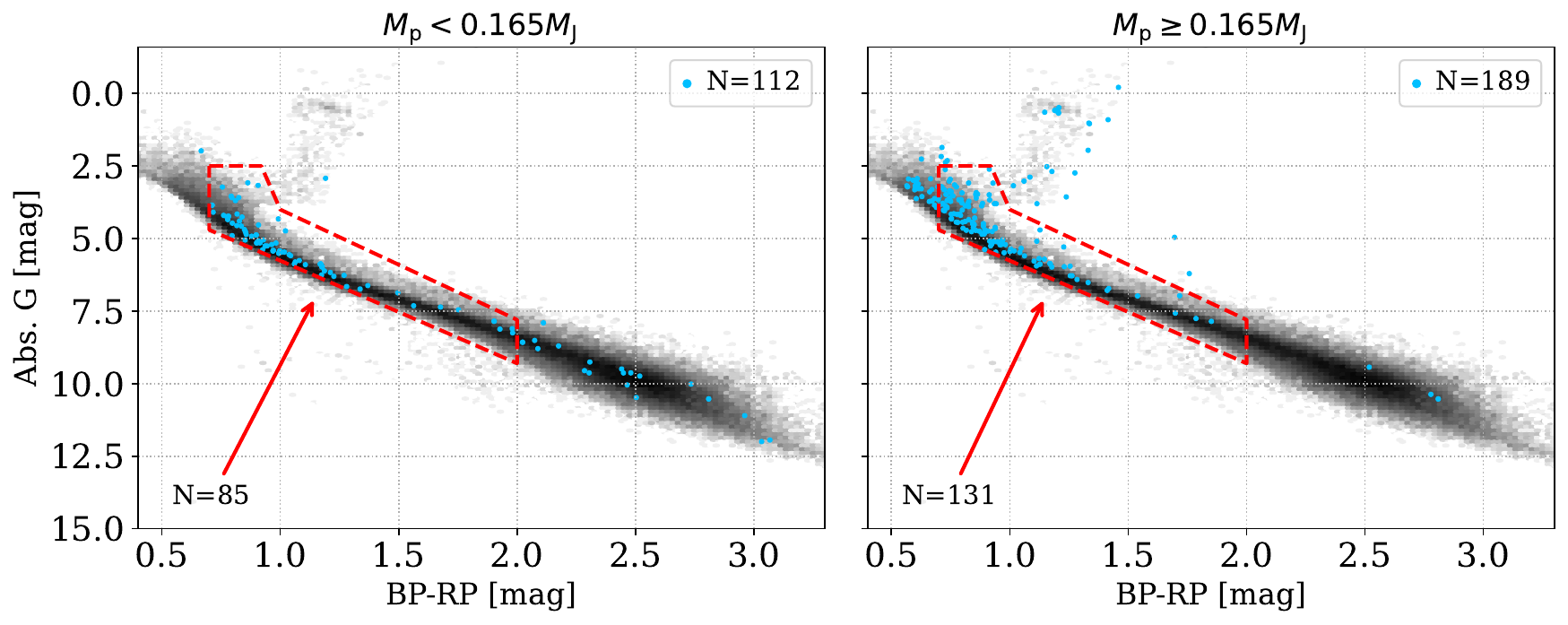}
\caption{Color-magnitude-diagram (CMD) of all planet-hosting stars with eccentric orbits in our sample. Left: low-mass planets, Right: high-mass 
planets. Background gray-scale sample is added for reference, using the Gaia Catalog of Nearby Stars (GCNS). Red polygons in both panels mark our selected sample of planet-host stars with similar stellar characteristics. See text for further description.}
\label{fig:cmd}
\end{figure*}

Next, we aimed to reduce possible biases of our analysis caused by the characteristics of the host stars, which might significantly influence the observed orbital parameters of their planets, either by inherent physical interactions or by some detection sensitivities \citep[see, for example][whose studies are based on the characteristics of the planet hosts]{banerjee24,Tang24}. 
To do so, Figure~\ref{fig:cmd} presents a color-magnitude-diagram (CMD) of the planet host-star samples, using \gaia G, and BP-RP magnitudes, separated to star hosting small and giant planets (blue points). The background gray-scale sample is added in both panels for reference and is based on the \gaia Catalog of Nearby Stars \citep[GCNS][]{GCNS}, a clean catalog of objects within $100$ pc of the Sun. As expected, 
the giant-planet hosts tend towards the early-type stars and evolved stars. To match the common region on the CMD where most small- and giant-planet host stars are situated, we selected only stars inside the red polygon\footnote{The polygon boundaries are as follows (BP-RP, Abs. G): (0.70, 2.5), (0.92, 2.5), (1, 4), (2, 7.8), (2, 9.3), (0.70, 4.7), (0.70, 2.5).} marked in Figure~\ref{fig:cmd}. We were left with a sample of $85$ small and $131$ giant planets.

These are the final two samples of eccentric planets we chose to analyze. 

\subsection{Discovery method}

While we exclusively focus on planetary orbits derived
from RV measurements, many of these planets were
initially identified through transit surveys before subsequent RV follow-ups. 
The specific statistical characteristics of transit-detected planets versus
those discovered through RV surveys vary based on the
distinct survey protocols and their respective detection
thresholds.
The transit detection method can
exhibit a bias toward planets with shorter
orbital periods, for example.
%

In our case, $49$ and $36$ of the small planets were first detected by the RV and transit methods, respectively, while in the case of the giant planets, the numbers are $64$ and $67$, respectively. In the next section, when exploring the PEP two distributions, we will also consider the discovery method of each planet, showing that the different methods have no impact on our results.

We list in tables \ref{tab:SmallPlanets} and \ref{tab:GiantPlanets} the full sample of low-mass and high-mass planet parameters of the orbital motion and host properties used in this work to assess the PEP distributions.

\begin{deluxetable*}{ccccccccc}[t] 
\tablecaption{Small Planets (\(M_{\rm p} < 0.165 M_{\rm J}\)) properties used in the analysis of the PEP distributions. The table includes properties of orbital period, eccentricity and lower eccentricity uncertainties, planet mass, number of other planets in the system, discovery method and \textit{Gaia} BP-RP and Abs. G host magnitudes. Sources are sorted by orbital period.
\label{tab:SmallPlanets}}
\tablewidth{0pt}
\tablehead{
\colhead{Name} & \colhead{Period} & \colhead{$e$} & \colhead{$\Delta e$} & \colhead{$M_{\rm p}$} & \colhead{Np} & \colhead{Discovery Method} & \colhead{BP-RP} & \colhead{Abs. G}\\
 & \colhead{[days]} &  & & \colhead{${\rm [} M_{\rm J}{\rm ]}$} & & & \colhead{[mag]} & \colhead{[mag]}
}
\startdata
GJ 2030 b& $2.3813 $ & $ 0.239$ & $ 0.058$ & $ 0.01$ & $ 2$ &RV & $ 0.91$ & $ 3.17$\\
TOI-1288 b& $2.6998 $ & $ 0.064$ & $ 0.015$ & $ 0.13$ & $ 2$ &Transit & $ 0.98$ & $ 5.15$\\
Kepler-411 b& $3.0052 $ & $ 0.146$ & $ 0.004$ & $ 0.08$ & $ 4$ &Transit & $ 1.23$ & $ 6.30$\\
HD 134060 b& $3.2696 $ & $ 0.450$ & $ 0.040$ & $ 0.03$ & $ 2$ &RV & $ 0.77$ & $ 4.25$\\
TOI-1272 b& $3.3160 $ & $ 0.338$ & $ 0.062$ & $ 0.08$ & $ 2$ &Transit & $ 1.11$ & $ 5.90$\\
GJ 676 A d& $3.6008 $ & $ 0.192$ & $ 0.079$ & $ 0.01$ & $ 4$ &RV & $ 1.90$ & $ 7.85$\\
HD 219828 b& $3.8348 $ & $ 0.091$ & $ 0.025$ & $ 0.06$ & $ 2$ &RV & $ 0.80$ & $ 3.59$\\
TOI-1062 b& $4.1130 $ & $ 0.177$ & $ 0.064$ & $ 0.03$ & $ 2$ &Transit & $ 1.00$ & $ 5.43$\\
HD 18599 b& $4.1374 $ & $ 0.340$ & $ 0.090$ & $ 0.08$ & $ 1$ &Transit & $ 1.06$ & $ 5.81$\\
HD 12235 b& $4.3631 $ & $ 0.376$ & $ 0.154$ & $ 0.03$ & $ 2$ &RV & $ 0.76$ & $ 3.22$\\
\enddata
\end{deluxetable*}

\begin{deluxetable*}{ccccccccc}[t] 
\tablecaption{Same as Table~\ref{tab:SmallPlanets} but for the giant planets sample (\(M_{\rm p} \geq 0.165 M_{\rm J}\)). \label{tab:GiantPlanets}}
\tablewidth{0pt}
\tablehead{
\colhead{Name} & \colhead{Period} & \colhead{$e$} & \colhead{$\Delta e$} & \colhead{$M_{\rm p}$} & \colhead{Np} & \colhead{Discovery Method} & \colhead{BP-RP} & \colhead{Abs. G}\\
 & \colhead{[days]} &  & & \colhead{${\rm [} M_{\rm J}{\rm ]}$} & & & \colhead{[mag]} & \colhead{[mag]}
}
\startdata
WASP-12 b& $1.0914 $ & $ 0.045$ & $ 0.004$ & $ 1.47$ & $ 1$ &Transit & $ 0.75$ & $ 3.42$\\
HAT-P-23 b& $1.2129 $ & $ 0.106$ & $ 0.044$ & $ 2.09$ & $ 1$ &Transit & $ 0.83$ & $ 4.35$\\
HAT-P-36 b& $1.3273 $ & $ 0.063$ & $ 0.023$ & $ 1.85$ & $ 1$ &Transit & $ 0.87$ & $ 4.75$\\
WASP-5 b& $1.6284 $ & $ 0.038$ & $ 0.018$ & $ 1.58$ & $ 1$ &Transit & $ 0.84$ & $ 4.51$\\
HD 86081 b& $2.1378 $ & $ 0.012$ & $ 0.005$ & $ 1.48$ & $ 1$ &RV & $ 0.76$ & $ 3.53$\\
WASP-140 b& $2.2360 $ & $ 0.047$ & $ 0.004$ & $ 2.44$ & $ 1$ &Transit & $ 0.98$ & $ 5.49$\\
HAT-P-16 b& $2.7760 $ & $ 0.036$ & $ 0.004$ & $ 4.19$ & $ 1$ &Transit & $ 0.72$ & $ 4.02$\\
HAT-P-20 b& $2.8753 $ & $ 0.015$ & $ 0.005$ & $ 7.25$ & $ 1$ &Transit & $ 1.42$ & $ 6.72$\\
HAT-P-13 b& $2.9162 $ & $ 0.013$ & $ 0.004$ & $ 0.85$ & $ 2$ &Transit & $ 0.86$ & $ 3.47$\\
HD 46375 b& $3.0236 $ & $ 0.063$ & $ 0.026$ & $ 0.23$ & $ 1$ &RV & $ 1.00$ & $ 5.36$\\
\enddata
\end{deluxetable*}
\section{PEP DISTRIBUTION FOR SMALL AND GIANT PLANETS}
\label{sec:PEP}

\begin{figure*}[t]
\includegraphics[width=18cm]{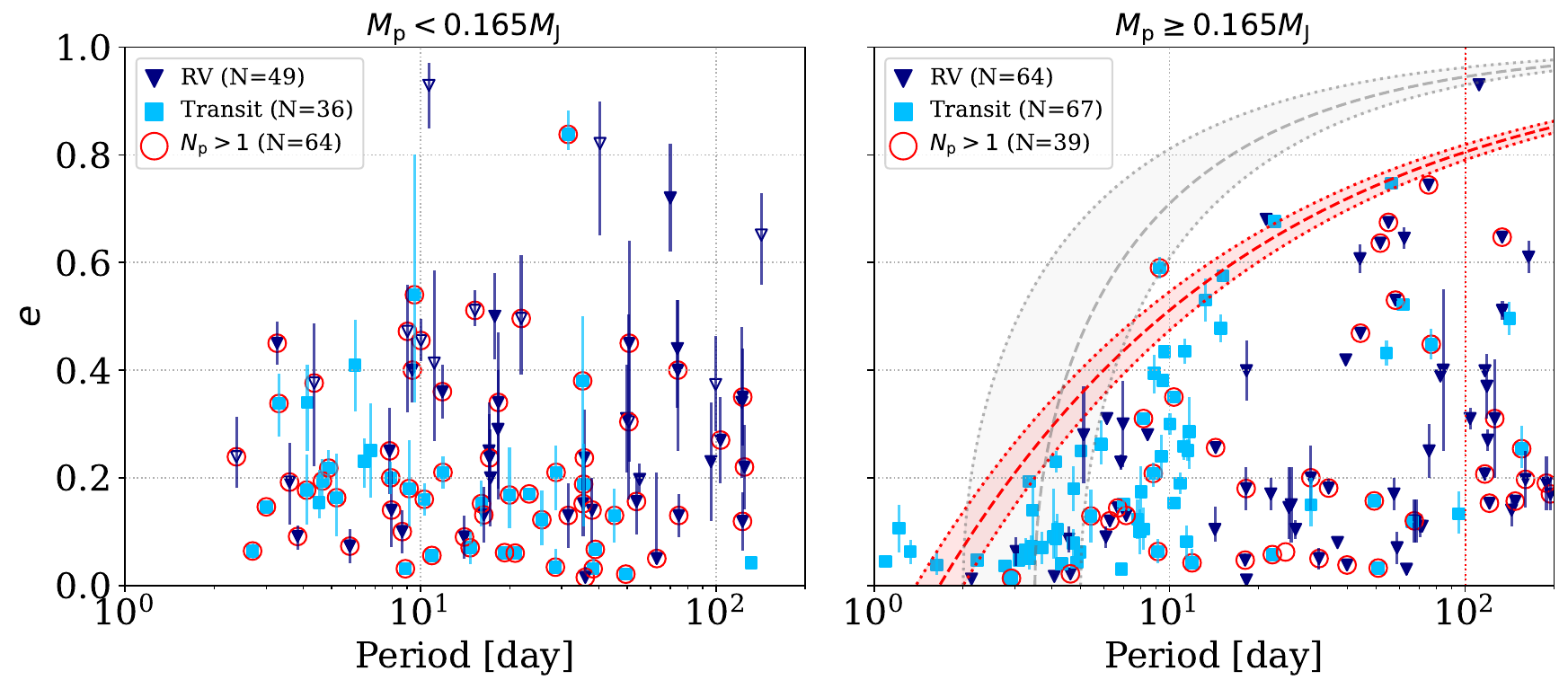}
\caption{Separate PEP diagrams for the small and giant planets. Triangles and squares represent planets discovered by the transit and RV methods, respectively. Red open circles mark planets in known multi-planet systems. The open triangles in the small-planet diagram present orbits from the study of \protect\cite{feng22}, which did not publish some of their orbital elements, nor the RV measurements. The red dashed line in the giant-planet diagram marks our upper-envelope best-fit model, marked by a vertical dotted line, based on planets with an orbital period shorter than $100$ days. The red area marks the transition region of $\pm \delta$ width along the envelope (see text).Gray dashed curve and the corresponding gray region mark the expected constant angular momentum evolutionary track, given a final orbital period of $2 \leq P_{\mathrm{final}} < 5$ days.}
    \label{fig:Per_Ecc_small_Giants}
\end{figure*}

Figure~\ref{fig:Per_Ecc_small_Giants} shows the PEP distributions of the low- and high-mass planets.

The difference between the two PEP diagrams is striking.
The high-mass orbits display a clear upper-envelope shape
that monotonically rises from zero eccentricity at the orbital period of $\sim 1.5$ days and reaches eccentricity of 0.8 at a period of $\sim 100$ days (see below). On the other hand, the low-mass planetary orbits display a flat eccentricity distribution between $\sim0.1$ and $\sim0.5$, with almost no dependence on the orbital period. 

To see if the pronounced contrast in the PEP between small and giant planets might be attributable to the detection method, we have segregated the data in Figure~\ref{fig:Per_Ecc_small_Giants} to distinguish between transit-detected planets (represented as squares) and those discovered through RV surveys (represented as triangles). No clear dependence on the detection method can be noticed (see below that the numerical fitting of the RV and transit detected planets of the giant planets yield similar upper envelopes). 

The low-mass PEP distribution includes five orbits with high eccentricities, clearly above the upper limit of the population at
$\sim 0.5$. 
Interestingly, all five systems have a distant sub-stellar companion, as detailed in Table~\ref{tab:ecc_small},
while the overall fraction of systems with known companions in the small-planet sample is $15$ out of $85$. This is based on the value marked by '\texttt{sy\_pnum}' and listed in the Exoplanet Archive
’Composite’ Planet Data Table, but note that the actual binary nature (separation, mass ratio) is not listed.
One might attribute these eccentricities to dynamical pumping induced by the distant companions 
\citep[e.g.,][see discussion]{mazeh79,fabrycky07}. 


\begin{deluxetable*}{lCCCCCCC}[h]
\tablecaption{Five inner small planets with high eccentricities and their distant companions. \label{tab:ecc_small}}
\tablewidth{0pt}
\tablehead{
\colhead{Name} & \multicolumn{3}{c}{Inner low-mass planet} & \multicolumn{3}{c}{Sub-stellar distant companion} & \colhead{Reference} \\
\cline{2-4} \cline{5-7}
\colhead{} & \colhead{Period } & \colhead{\(e\)} & \colhead{$M_{\rm p}$} & \colhead{Period } & \colhead{\(e\)} & \colhead{$M_{\rm p}$} & \colhead{}\\
& \colhead{[days]}&  &   \colhead{ \([M_{\rm J}]\)} & \colhead{[days]} &  &  \colhead{ \([M_{\rm J}]\)} & \colhead{}
}
\startdata
GJ-3222 & 10.667 & 0.93\pm0.06 & 0.036\pm0.008 & 1429 & 0.537\pm0.052 & 52\pm5 & [1] \\
Kepler-1656 & 31.562 & 0.838\pm0.33 & 0.15 \pm 0.02 & 1919 & 0.527\pm0.051& 0.40\pm 0.09 & [2] \\
HD 144899 & 40.439 & 0.82\pm0.15 & 0.064\pm0.018 & 8113 &0.355\pm0.019 &100\pm10 & [1] \\
GJ 2056 & 69.971 & 0.72\pm 0.10 & 0.051\pm0.011 &2982 & 0.81\pm 0.018&0.44\pm0.05 & [1] \\
HD 13724 & 142.228 & 0.651\pm0.083 & 0.16\pm0.023 & 15784 & 0.327\pm0.029&36\pm2 & [1] \\
\enddata
\tablecomments{References: [1] \cite{feng22}; [2] \cite{Angelo22}. }
\end{deluxetable*}

We mark by red circle in the left panel of the low-mass planets sample of Figure~\ref{fig:Per_Ecc_small_Giants} the sources that are part of known multi-planet systems,  $N_{\mathrm{p}}>1$, based on the value marked by '\texttt{sy\_pnum}' and listed in the Exoplanet Archive ’Composite’ Planet
Data Table. The majority of planets in the low-mass sample ($64$ out of $85$) have at least one additional detected planet in the same stellar system. In contrast, for the high-mass planet sample, the fraction of ones with other planets orbiting the same host star is much smaller ($39$ out of $131$). 

To quantify the observed difference in the two PEP diagrams, we used a KS-test to compare the eccentricity distributions of planets with orbital periods shorter and longer than $10$ days. The p-values obtained were $0.88$ and $9.2 \times 10 ^{-4}$ for the low- and high-mass planetary samples respectively. These results further confirm the observed trend: while the eccentricity distributions of low-mass planets remain consistent across the orbital period, those of high-mass planets exhibit significant variations depending on the period cut. This suggests a distinct dependency of eccentricity on orbital period for planets of higher mass.

In the following we fit an upper envelope to the giant-planet PEP diagram.

\subsection{Modeling the upper envelope} 

To numerically characterize the upper envelope seen in the observed high-mass PEP, we apply the methodology constructed by \cite*{Bashi23},
which defined a functional form of two parameters: 
%
\begin{equation}
f(P) = \left\{
\begin{aligned}
&1-\left(\frac{P_{\mathrm{cut}}}{P}\right)^{\frac{1}{\tau}}, && \text{if } P > P_{\mathrm{cut}}\ , \\
&0, && \text{if } P \leq P_{\mathrm{cut}}\ ,
\end{aligned}
\right.
\label{eq:e_p_env}
\end{equation}
%
%
where \Pcut\ is the cutoff period and $\tau$ is a dimensionless parameter that determines the slope of the envelope. The function gets a zero value at \Pcut\  for any value of $\tau$, and asymptotically rises to unity as $P$ gets substantially longer than \Pcut. 

We describe the probability density function (PDF) for planets in the ($\log P$, $e$) plane using a modified Fermi function, $\mathcal{F}_{\mathrm{PDF}}(P,e;\,{P_{\mathrm{cut}}},\tau,\delta)$. The PDF converges to zero above the envelope, and to a positive constant below it, with a transition region of width $\delta$ along the envelope.

For each set of parameters of the PDF---${P}_{\rm cut}$, $\tau$ and $\delta$,
we derive the likelihood of the sample as
\begin{equation}
\mathcal{L} = \prod_{i=1}^{N} \mathcal{F}_{\mathrm{PDF}}({P}_i, e_i; {P}_{\rm cut},\tau,\delta) \ .
\label{eq:likelihood}
\end{equation}
%
where $e_i$ and $P_i$ are the eccentricity and period of the $i-$th planet. We search the three parameters that maximize the likelihood of the observed sample.

\subsection{The upper Envelope of the high-mass PEP diagram}

We applied our algorithm to the PEP of the high-mass planets sample, using an MCMC routine with $50$ walkers and $10^4$ steps, with uninformative priors for the three parameters.
We used the Python \texttt{emcee} package \citep{foreman-Mackey2013} to find the best parameter values and their uncertainties. Our prior boundaries and best-fit posterior values are given in Table~\ref{tab:params}. The red dashed line in the high-mass diagram of Figure~\ref{fig:Per_Ecc_small_Giants} shows our best-fit upper envelope. The red area in the figure marks the transition region of $\pm \delta$ width along the upper envelope.

A single outlier in the figure, somewhat above our fitted envelope, is that of CoRoT-20~b \citep{Deleuil12} with $e=0.59\pm 0.02$ and $P= 9.2429\pm 0.0003$ days. Interestingly, in a later work, \cite{Rey18} have found this system hosts another sub-stellar eccentric companion, CoRoT-20~c, with $M_{\mathrm{p}} \sin i=17 \pm 1 M_{\mathrm{J}}$, $e= 0.60 \pm 0.03$ and $P= 4.59\pm 0.05$ yr, suggesting a dynamical eccentricity pumping induced by the distant companion (see above). 

In addition, there are three outliers located left to our upper envelope fit with orbital periods shorter than $2$ days and significant eccentricities. These planets are WASP-12~b \citep[$e= 0.0447 \pm 0.0045$,][]{Hebb09}, HAT-P-36~b \citep[$e= 0.063 \pm 0.023$,][]{Wang19} and HAT-P-23~b \citep[$e= 0.106 \pm 0.044$,][]{Bakos21}. 
%
%
The small but significant eccentricities for planets so close to their host stars are unexpected and might be interesting candidates for further exploration. We suggest, as discussed in \cite{Arras12} and more recently pointed out by \cite{YeeWinn20} in the case of WASP-12~b, that these eccentricities are likely artifacts of the tidal interactions induced by the planet on its host rather than true orbital characteristics. In their work, \cite{Arras12} have demonstrated how the tidal fluid flow and the epicyclic motion of a slightly eccentric orbit produce an RV signal at half the orbital period. These signals can consequently mimic the effects of a non-zero eccentricity, leading to the erroneous conclusion that the orbit is eccentric when it is, in fact, circular. 

We further note, that these three planets were detected as part of large ground-based transit search surveys: HAT \citep{HAT04} and WASP \citep{WASP06}. Given the significantly larger number of stars these surveys can monitor compared to radial velocity (RV) surveys, we find it reasonable that the shortest orbital period giant planets were discovered first by transiting surveys where the transit detection probability is also the highest given the geometric probability.

\begin{deluxetable}{lll}[t]
\tablecaption{Prior and posterior values of upper envelope model of the high-mass planet sample.\label{tab:params}}
\tablewidth{0pt}
\tablehead{
\colhead{Parameter} & \colhead{Prior} & \colhead{Posterior}
}
\startdata
\( P_{\mathrm{cut}}\) & \(\log \mathcal{U}(-1,1)\) & \(1.66^{+0.24}_{-0.25}~\mathrm{day}\) \\
\(\tau\) & \(\log \mathcal{U}(0,2)\) & \(2.50^{+0.29}_{-0.22}\) \\
\(\delta\) & \(\mathcal{U}(0.01,0.3)\) & \(0.079^{+0.032}_{-0.023}\) \\
\enddata
\end{deluxetable}


To look for potential bias in the derived envelope due to the planetary discovery method, we conducted separate fittings of upper envelopes for the subsets of high-mass planets identified by the transit and RV methods. 
We found the upper-envelope parameters of the two sub-samples to be consistent with each other. 
The  $\log P_{cut}$ values were found to be $0.15^{+0.08}_{-0.12}$ and $0.35^{+0.17}_{-0.12}$ for the transit and RV detected sub-samples, respectively, indicating $\sim1.5\sigma$ difference, probably due to the bias of transit detections towards short periods.
For the slope of the envelope, $\log \tau$, we found 
$0.94^{+0.19}_{-0.14}$ and
$1.21^{+0.26}_{-0.15}$, respectively, within $\sim1\sigma$ of each other. 

To further explore the impact of the uncertainties of the orbital parameters on our upper envelope posterior estimates, we employed a bootstrap approach. We randomly drew eccentricities for each planet in the sample, using a normal distribution with the mean and standard deviation corresponding to that planet's reported mean and uncertainty values, as listed in the Exoplanet Archive 'Composite' Planet Data Table. In this way, we generated revised orbital parameters for our 131 high-mass planets and then applied the algorithm of finding the parameters of the upper envelope.
By this way we got distributions of the values of the three parameters, based on the eccentricity uncertainties. 
Overall, we found that the bootstrap uncertainties for the upper envelope posterior values were an order of magnitude smaller than the estimated uncertainties given by the final MCMC model. This finding affirms that our results are insensitive to the uncertainties of the planetary orbital eccentricities.

\section {Discussion}        %
\label{sec:discussion}       %

This study considered $1040$ planetary orbits with periods in the $1 \leq P < 200$ days range with RV solutions, as appeared in the Exoplanet Archive ’Composite’ catalog database.
The sample was not obtained by any systematic search but is a compilation of results reported by many studies. Therefore, it is subject to different observational detection thresholds and to the ability of the different projects to derive the orbital eccentricity correctly. Nevertheless, the trends found are highly significant and do not depend on the completeness of the sample. 



Of the 1040 orbits, only 301 have significant eccentricity, with $e >2\Delta e$ that can be used to explore the PEP distribution. 
After considering possible biases in detection method and host properties, we were left with a sample of $131$  high-mass orbits. This sample 
displays a clear upper envelope, which we model using our novel formalism \citep*{Bashi23}. We find that the upper envelope
monotonically rises from zero eccentricity at an orbital period of $\sim 1.7$ days and reaches an eccentricity of 0.8 at a period of $100$ days. 
The $85$ low-mass planetary orbits, on the other hand, display a flat eccentricity distribution between $\sim0.1$ and $\sim0.5$, with almost no dependence on the orbital period. As seen from the diagram and supported by our separate analysis, these results are independent of the planetary discovery method.


\subsection{Tidal Circularization---small versus giant planets}

The giant PEP distribution
has a typical upper-envelope signature of tidal circularization, similar to the eccentricity distribution of spectroscopic binaries, as analyzed recently by \cite*{Bashi23}, and was interpreted as the result of circularization, either during the stellar lifetime or at the pre-main-sequence stellar phase. 
The small-planet PEP distribution, on the other hand, is very far from having the shape of a sample that went through tidal circularization. 
Can we explain the substantially different PEP distributions?


We start by comparing the circularization timescale $\tau_{\rm circ}$ of the small- and giant-planets, using  a formula derived from \cite{GoldreichSoter66} theory: 
\begin{equation}
\tau_{\rm circ} = \left(4/63\right) Q'_{\rm p} 
\frac{M_{\rm p}}{M_*} \left( \frac{a}{R_{\rm p}} \right)^5 
P_{\rm orb}
\, ,
  \label{eq:t_circ}
\end{equation}
%
where $a$ is the orbital semi-major axis, $M_*$ is the stellar mass, and $M_{\rm p}$, $R_{\rm p}$ and $P_{\rm orb}$ are the planet mass, radius and orbital period, respectively. 
The tidal quality factor $Q'_{\rm p}$ is a dimensionless parameter that quantifies the efficiency of energy dissipation due to tidal forces \citep{GoldreichSoter66}. 

Consider, for example, an orbit of $4$ days, for which $a\sim 10 R_{\odot}$. For a giant planet, with $M_p/M_{\odot}\sim10^{-3}$ and $R_p\sim10^{-1}R_{\odot}$ and a solar-type star one gets  
%
\begin{equation} 
{\rm Giant\, Planets:} \,\,  \frac{M_{\rm p}}{M_*} \left( \frac{a}{R_{\rm p}} \right)^5 \sim 10^{7}
 \, ,
  \label{eq:giant}
\end{equation}
%
while for small-mass planets,
$M_p/M_{\odot}\sim3 \times\,10^{-5}$ and $R_p\sim10^{-2}R_{\odot}$ \citep[e.g.,][]{ChabrierBaraffe09,Bashi17, Ulmer-Moll19}, and therefore
%
\begin{equation} 
{\rm Small\, Planets:} \,\,  \frac{M_{\rm p}}{M_*} \left( \frac{a}{R_{\rm p}} \right)^5 \sim 3\, \times 10^{10}
 \, . 
  \label{eq:small}
\end{equation}

Previous works have argued that the typical quality factor of giant planets is $Q_{\rm p}\sim 10^5$, while rocky planets of the solar systems usually have $Q_{\rm p}\sim 10^2$ \citep{GoldreichSoter66, MurrayDermott99, ClausenTilgner15}.
We adopt these values, although they represent order-of-magnitude estimation only \citep[e.g.,][]{MatsumaraTakedaRasio09,papaloizou11}. 
When we put the corresponding values in Equation~\ref{eq:t_circ} we get 
%
\begin{equation} 
{\rm Giant\, Planets:} \,\,  
\tau_{\rm circ,g}\sim 10^{11}P_{\rm orb}\sim 10^9 {\rm yrs}
 \, ,
  \label{eq:T_circ_giant}
\end{equation}
%
while for small-mass planets, 
%
\begin{equation} 
{\rm Small\, Planets:} \,\,  
\tau_{\rm circ,s}\sim 3\times\,10^{11}P_{\rm orb}\sim 3\times\,10^9 {\rm yrs}
 \, . 
  \label{eq:T_circ_small}
\end{equation}
%
Note that the ratio between the two timescales, 
$\tau_{\rm circ,s}/\tau_{\rm circ,g}\sim3$ ,  
is independent of the orbital period. The longer circularization timescale of the small planets indicates a weaker tidal interaction relative to that of the giant planets by a factor of $\sim 3$, all other binary elements being equal. The two timescales are order-of-magnitude estimations only, because of the spread of periods and masses of the two populations.  

The small factor difference between the estimations of the two types of planets cannot explain the striking difference between the two PEP distributions. 
If the small planets went through a similar tidal circularization, 
we could expect their PEP distribution to show a similar upper-envelope shape, with a slightly longer $P_{\rm cut}$ that compensates for the factor $3$ weakness of their tidal interaction. Apparently, the small planets were not able to go through tidal circularization. 

On the other hand, if the dissipation is dominated by friction inside the parent star, it would provide an easy explanation of the difference we have observed between the PEP diagrams of low-mass and high-mass planets.  Obviously, low-mass planets do not perturb the star as much as high-mass planets.

As suggested by previous studies \citep[e.g.,][]{Eylen19}, the eccentricity distribution of the small planets might be due to the fact that many of the small planets were formed with neighboring planets, probably at relatively large distances from their parent star. Supporting this statement is the fact that most of the close-in low-mass planets in our sample reside in multi-planet systems, as suggested by Figure~\ref{fig:Per_Ecc_small_Giants} \citep[see, for example][]{murray23}.
These planets then migrated inwards, while probably dynamically locked with their neighbouring planets \citep[see, for example, the discussion by][]{weiss23}.
%
These "multis" were under strong gravitational interaction with their neighbors, which could inflate their eccentricities. On the other hand, eccentricities too high, with, say, $e >0,5$, in one of the orbits, can render the configuration dynamically unstable
\citep[e.g.,][]{tremaine15,correia20,goldberg22,murray23,ghosh23}; see also \cite{matsumoto15}.

Note, however, that not all small planets are known to reside in planetary multiple systems. This could be because some of the neighboring planets are undetectable, as their mass, inclination, and/or period push them below the observational RV detection threshold. The eccentricity distribution as a function of the multiplicity of the planetary systems \citep[e.g.,][]{EylenAlbrecht15,Eylen19,ballard23} could be key evidence for understanding the role of the neighboring planets in planetary eccentricity. This is however behind the scope of this paper.  

As a final remark, we note that data irregularities and sparse observational coverage can occasionally lead to inaccurate interpretations of orbital configurations. For instance, what appears to be a single low-mass, highly eccentric exoplanet could be two separate planets each in a more typical, near-circular orbit \citep{Wittenmyer19}. This kind of misinterpretation can occur when signals from two closely spaced or overlapping planets are not sufficiently sampled, causing them to be mistaken for a single eccentric orbit.

\subsection{Tidal Circularization--- giant planets versus binaries}


Interestingly, the $P_{\rm cut}$ of the giant-planet distribution, $\sim 1.8$ day, is significantly shorter than the one obtained for the spectroscopic binaries, which is about $\sim 6$ days for similar primary stars. 
We note that if we apply Equation~\ref{eq:t_circ} to binaries with stellar companions, replacing the planetary mass and radius by stellar mass and radius, $M_{\rm comp}$ and $R_{\rm comp}$, we get, assuming the same $Q$, 
%
\begin{equation}
\tau_{\rm circ}
\propto
\frac{M_{\rm comp}} {R_{\rm comp}^5}
\, .
  \label{eq:t_circII}
\end{equation}
%
This leads to an estimation of the stellar circularization timescale, for solar mass $M_{\rm comp}\sim 10^3 M_{\rm J}$ and radius $R_{\rm comp}\sim 10 R_{\rm J}$ , 
%
\begin{equation} 
\tau_{\rm circ,stellar}   \sim 10^{-2} \tau_{\rm circ,g} 
 \, .
  \label{eq:T_circ_stellar}
\end{equation}

This implies that the circularization processes are more effective in binaries, and therefore could circularize systems with longer periods, 
if they were active during the same period of time, the stellar lifetime, for example. This is consistent with what has been observed.


\subsection{Periastron period}
As the circularization is most effective during periastron passage, we might expect that the upper envelope represents a constant periastron distance. Planets with too short periastron passage went through fast tidal interaction. This means that for any given period the eccentricity has a maximum value, parameterized by  $P_{\mathrm{peri}}$, such that  
%
\begin{equation}
e_{\rm max} = 1 - \left( \frac{P_{\mathrm{peri}}}{P} \right )^{2/3}
\, .
  \label{eq:peri_per}
\end{equation}
%
This form is the same as our upper-envelope function,
which is translated to
 $1/\tau \sim 2/3$. 
 Our finding of $1/\tau \sim 0.4$, a value not far from the one expected, given the scatter of the sample and the simplifying assumptions.

\subsection{Exploring the constant angular momentum track}
As a final test, we wish to explore the expected 'evolutionary track' of HEM giant planets assuming they conserve their angular momentum. Given the period distribution of circular hot Jupiters has a strong 3-day accumulation \citep[e.g.][]{Cumming99, Udry03, Cumming08}, we chose to use this circularized 'final' orbital period, $P_{\mathrm{final}}$ to rewind the hot Jupiter’s tidal evolution history using \citep{dawson18,Jackson23}
%
\begin{equation}
P_{\mathrm{final}}=P(t)\left(1-e^2 \right)^{3/2}
\, .
  \label{eq:hem}
\end{equation}
%
We plot in the right panel of Figure~\ref{fig:Per_Ecc_small_Giants} 
the expected 'evolutionary track' of these objects given we find most circularized hot-Jupiters in the 
$2 \leq P_{\mathrm{final}} < 5$ days limits.
Clearly, these evolutionary tracks are different from the derived upper envelope of the sample.

\subsection{Highly eccentric planetary orbits}

Five orbits of small planets do show high eccentricities larger than $0.6$. Interestingly, all five systems are known to have a sub-stellar companion at an orbital period of $1000$--$16,000$ days. Those companions might be responsible for pump-in high eccentricity into the planetary orbit during the formation or evolution of these planets  
\citep[e.g.,][]{fabrycky07, correia12, petrovich14}. 
Obviously, one needs a systematic search for distant companions of all small planets, not only the ones with high eccentricities, to assess the contribution of the assumed eccentricity pumping.  

In fact, there are a few known {\it giant} planets with very eccentric orbits, with periods longer than the range considered here. One of the most well-known systems is that of HD 80606b (also seen in our PEP diagram) with a period of $\sim 111$ days and $e=0.927 \pm 0.012$ \citep{Naef01}.
HD 20782b is another example, with a period of $\sim 600$ days and $e=0.97 \pm 0.01$ \citep{otoole09}.
The most recently discovered one, HIP 66074b, 
with a period of $\sim 300$ days and $e=0.948 \pm 0.004$,
has a distant companion that might also have a sub-stellar mass \citep[see discussion by][]{sozzetti23}.
For these cases, the periastron distance, $\sim 0.05$ au, corresponds to about the histogram peak of circular orbits of giant planets at $\sim 4$ days.

%

\section{Conclusion}
\label{sec:conclusions}
The present PEP probably reflects both the primordial eccentricity and the circularization processes. 
The peak of the period distribution of the circular orbits at $3$--$4$ days apparently presents the giant planets whose orbits have already been circularized, as suggested by the 
the HEM model \citep{Socrates12, Bonomo17, konigel17, dawson18, Jackson23}.
To find the primordial periods and eccentricities of these orbits, one needs to integrate back the orbits with the equations that govern the eccentricity and period evolution. 
Note, however, that a backward model might be quite complex, as it might need to include a possible orbital misalignment of the spin axes of the parent star and the planet \citep[e.g.,][]{MardlingLin02, vick23}, unseen small planets, and differential stellar rotation, which could complicate the model. 
 


Finally, it is interesting to note that the eccentricity of the exoplanet population revealed in the last three decades surprising features \cite[see extensive discussion in][]{biazzo22}, in contrast to the Solar system, where all planets have almost circular orbits. 
The dependence of the eccentricity on the planetary mass and orbital period was studied here, suggesting that the giant planets went through tidal circularization processes while the small planets did not.
Other features, like the 
dependence of the eccentricity on the parent star metallicity \citep{An23} and the planetary multiplicity 
\citep{EylenWinnAlbrecht16}, are also extremely important. All these features
carry with them many hints on the formation and evolution of the exoplanet population.

\section{Acknowledgments}
We wish to thank the anonymous reviewer for an enlightening report. We extend our heartfelt gratitude to Josh N. Winn and Daniel C. Fabrycky for adding further comments and suggestions on an early version of this work.  This research was supported by Grant No. 2016069 of the United States-Israel Binational Science Foundation (BSF) to TM, 
and
Grant No.~I-1498-303.7/2019 of the German-Israeli Foundation for Scientific Research and Development (GIF) to TM. This research is partially supported by Israel Ministry of Innovation, Science and Technology grant 0004138. 
D.B. acknowledges the support of the Blavatnik family and
the British council as part of the Blavatnik Cambridge Fellowship.
This research has made use of the NASA Exoplanet Archive, which is operated by the California Institute of Technology, under contract with the National Aeronautics and Space Administration under the Exoplanet Exploration Program.
This work has also made use of data from the European Space Agency (ESA) mission \textit{Gaia} (https://www.cosmos.esa.int/gaia), processed by the \textit{Gaia} Data Processing and Analysis Consortium (DPAC; https://www.cosmos.esa.int/web/gaia/
dpac/consortium). Funding for DPAC has been provided by national institutions, in particular the institutions participating in the \textit{Gaia} Multilateral Agreement.

%
\facility{Exoplanet Archive}
\vspace{5mm}


\software{astropy \citep{Astropy13,Astropy18}, NumPy \citep{NumPy}, Matplotlib \citep{Matplotlib}, scikit-learn \citep{pedregosa2011scikit}, SciPy \citep{SciPy}, emcee \citep{foreman-Mackey2013}
          }


\bibliography{main_arxiv}{}

\begin{thebibliography}{}
\expandafter\ifx\csname natexlab\endcsname\relax\def\natexlab#1{#1}\fi
\providecommand{\url}[1]{\href{#1}{#1}}
\providecommand{\dodoi}[1]{doi:~\href{http://doi.org/#1}{\nolinkurl{#1}}}
\providecommand{\doeprint}[1]{\href{http://ascl.net/#1}{\nolinkurl{http://ascl.net/#1}}}
\providecommand{\doarXiv}[1]{\href{https://arxiv.org/abs/#1}{\nolinkurl{https://arxiv.org/abs/#1}}}

\bibitem[{{Adibekyan}(2019)}]{Adibekyan19}
{Adibekyan}, V. 2019, Geosciences, 9, 105, \dodoi{10.3390/geosciences9030105}

\bibitem[{{Akeson} {et~al.}(2013){Akeson}, {Chen}, {Ciardi}, {Crane}, {Good}, {Harbut}, {Jackson}, {Kane}, {Laity}, {Leifer}, {Lynn}, {McElroy}, {Papin}, {Plavchan}, {Ram{\'\i}rez}, {Rey}, {von Braun}, {Wittman}, {Abajian}, {Ali}, {Beichman}, {Beekley}, {Berriman}, {Berukoff}, {Bryden}, {Chan}, {Groom}, {Lau}, {Payne}, {Regelson}, {Saucedo}, {Schmitz}, {Stauffer}, {Wyatt}, \& {Zhang}}]{Akeson13}
{Akeson}, R.~L., {Chen}, X., {Ciardi}, D., {et~al.} 2013, \pasp, 125, 989, \dodoi{10.1086/672273}

\bibitem[{{An} {et~al.}(2023){An}, {Xie}, {Dai}, \& {Zhou}}]{An23}
{An}, D.-S., {Xie}, J.-W., {Dai}, Y.-Z., \& {Zhou}, J.-L. 2023, \aj, 165, 125, \dodoi{10.3847/1538-3881/acb533}

\bibitem[{{Angelo} {et~al.}(2022){Angelo}, {Naoz}, {Petigura}, {MacDougall}, {Stephan}, {Isaacson}, \& {Howard}}]{Angelo22}
{Angelo}, I., {Naoz}, S., {Petigura}, E., {et~al.} 2022, \aj, 163, 227, \dodoi{10.3847/1538-3881/ac6094}

\bibitem[{{Arras} {et~al.}(2012){Arras}, {Burkart}, {Quataert}, \& {Weinberg}}]{Arras12}
{Arras}, P., {Burkart}, J., {Quataert}, E., \& {Weinberg}, N.~N. 2012, \mnras, 422, 1761, \dodoi{10.1111/j.1365-2966.2012.20756.x}

\bibitem[{{Astropy Collaboration} {et~al.}(2013){Astropy Collaboration}, {Robitaille}, {Tollerud}, {Greenfield}, {Droettboom}, {Bray}, {Aldcroft}, {Davis}, {Ginsburg}, {Price-Whelan}, {Kerzendorf}, {Conley}, {Crighton}, {Barbary}, {Muna}, {Ferguson}, {Grollier}, {Parikh}, {Nair}, {Unther}, {Deil}, {Woillez}, {Conseil}, {Kramer}, {Turner}, {Singer}, {Fox}, {Weaver}, {Zabalza}, {Edwards}, {Azalee Bostroem}, {Burke}, {Casey}, {Crawford}, {Dencheva}, {Ely}, {Jenness}, {Labrie}, {Lim}, {Pierfederici}, {Pontzen}, {Ptak}, {Refsdal}, {Servillat}, \& {Streicher}}]{Astropy13}
{Astropy Collaboration}, {Robitaille}, T.~P., {Tollerud}, E.~J., {et~al.} 2013, \aap, 558, A33, \dodoi{10.1051/0004-6361/201322068}

\bibitem[{{Astropy Collaboration} {et~al.}(2018){Astropy Collaboration}, {Price-Whelan}, {Sip{\H{o}}cz}, {G{\"u}nther}, {Lim}, {Crawford}, {Conseil}, {Shupe}, {Craig}, {Dencheva}, {Ginsburg}, {Vand erPlas}, {Bradley}, {P{\'e}rez-Su{\'a}rez}, {de Val-Borro}, {Aldcroft}, {Cruz}, {Robitaille}, {Tollerud}, {Ardelean}, {Babej}, {Bach}, {Bachetti}, {Bakanov}, {Bamford}, {Barentsen}, {Barmby}, {Baumbach}, {Berry}, {Biscani}, {Boquien}, {Bostroem}, {Bouma}, {Brammer}, {Bray}, {Breytenbach}, {Buddelmeijer}, {Burke}, {Calderone}, {Cano Rodr{\'\i}guez}, {Cara}, {Cardoso}, {Cheedella}, {Copin}, {Corrales}, {Crichton}, {D'Avella}, {Deil}, {Depagne}, {Dietrich}, {Donath}, {Droettboom}, {Earl}, {Erben}, {Fabbro}, {Ferreira}, {Finethy}, {Fox}, {Garrison}, {Gibbons}, {Goldstein}, {Gommers}, {Greco}, {Greenfield}, {Groener}, {Grollier}, {Hagen}, {Hirst}, {Homeier}, {Horton}, {Hosseinzadeh}, {Hu}, {Hunkeler}, {Ivezi{\'c}}, {Jain}, {Jenness}, {Kanarek}, {Kendrew}, {Kern}, {Kerzendorf}, {Khvalko}, {King}, {Kirkby}, {Kulkarni},
  {Kumar}, {Lee}, {Lenz}, {Littlefair}, {Ma}, {Macleod}, {Mastropietro}, {McCully}, {Montagnac}, {Morris}, {Mueller}, {Mumford}, {Muna}, {Murphy}, {Nelson}, {Nguyen}, {Ninan}, {N{\"o}the}, {Ogaz}, {Oh}, {Parejko}, {Parley}, {Pascual}, {Patil}, {Patil}, {Plunkett}, {Prochaska}, {Rastogi}, {Reddy Janga}, {Sabater}, {Sakurikar}, {Seifert}, {Sherbert}, {Sherwood-Taylor}, {Shih}, {Sick}, {Silbiger}, {Singanamalla}, {Singer}, {Sladen}, {Sooley}, {Sornarajah}, {Streicher}, {Teuben}, {Thomas}, {Tremblay}, {Turner}, {Terr{\'o}n}, {van Kerkwijk}, {de la Vega}, {Watkins}, {Weaver}, {Whitmore}, {Woillez}, {Zabalza}, \& {Astropy Contributors}}]{Astropy18}
{Astropy Collaboration}, {Price-Whelan}, A.~M., {Sip{\H{o}}cz}, B.~M., {et~al.} 2018, \aj, 156, 123, \dodoi{10.3847/1538-3881/aabc4f}

\bibitem[{{Bakos} {et~al.}(2004){Bakos}, {Noyes}, {Kov{\'a}cs}, {Stanek}, {Sasselov}, \& {Domsa}}]{HAT04}
{Bakos}, G., {Noyes}, R.~W., {Kov{\'a}cs}, G., {et~al.} 2004, \pasp, 116, 266, \dodoi{10.1086/382735}

\bibitem[{{Bakos} {et~al.}(2011){Bakos}, {Hartman}, {Torres}, {Latham}, {Kov{\'a}cs}, {Noyes}, {Fischer}, {Johnson}, {Marcy}, {Howard}, {Kipping}, {Esquerdo}, {Shporer}, {B{\'e}ky}, {Buchhave}, {Perumpilly}, {Everett}, {Sasselov}, {Stefanik}, {L{\'a}z{\'a}r}, {Papp}, \& {S{\'a}ri}}]{Bakos21}
{Bakos}, G.~{\'A}., {Hartman}, J., {Torres}, G., {et~al.} 2011, \apj, 742, 116, \dodoi{10.1088/0004-637X/742/2/116}

\bibitem[{{Banerjee} {et~al.}(2024){Banerjee}, {Narang}, {Manoj}, {Henning}, {Tyagi}, {Surya}, {Nayak}, \& {Tripathi}}]{banerjee24}
{Banerjee}, B., {Narang}, M., {Manoj}, P., {et~al.} 2024, arXiv e-prints, arXiv:2404.16499, \dodoi{10.48550/arXiv.2404.16499}

\bibitem[{{Bashi} {et~al.}(2018){Bashi}, {Helled}, \& {Zucker}}]{Bashi18}
{Bashi}, D., {Helled}, R., \& {Zucker}, S. 2018, Geosciences, 8, 325, \dodoi{10.3390/geosciences8090325}

\bibitem[{{Bashi} {et~al.}(2017){Bashi}, {Helled}, {Zucker}, \& {Mordasini}}]{Bashi17}
{Bashi}, D., {Helled}, R., {Zucker}, S., \& {Mordasini}, C. 2017, \aap, 604, A83, \dodoi{10.1051/0004-6361/201629922}

\bibitem[{{Bashi} {et~al.}(2023){Bashi}, {Mazeh}, \& {Faigler}}]{Bashi23}
{Bashi}, D., {Mazeh}, T., \& {Faigler}, S. 2023, \mnras, 522, 1184, \dodoi{10.1093/mnras/stad999}

\bibitem[{{Biazzo} {et~al.}(2022){Biazzo}, {Bozza}, {Mancini}, \& {Sozzetti}}]{biazzo22}
{Biazzo}, K., {Bozza}, V., {Mancini}, L., \& {Sozzetti}, A. 2022, in Astrophysics and Space Science Library, Vol. 466, Demographics of Exoplanetary Systems, Lecture Notes of the 3rd Advanced School on Exoplanetary Science, ed. K.~{Biazzo}, V.~{Bozza}, L.~{Mancini}, \& A.~{Sozzetti}, 143--234, \dodoi{10.1007/978-3-030-88124-5_3}

\bibitem[{{Bonomo} {et~al.}(2017){Bonomo}, {Desidera}, {Benatti}, {Borsa}, {Crespi}, {Damasso}, {Lanza}, {Sozzetti}, {Lodato}, {Marzari}, {Boccato}, {Claudi}, {Cosentino}, {Covino}, {Gratton}, {Maggio}, {Micela}, {Molinari}, {Pagano}, {Piotto}, {Poretti}, {Smareglia}, {Affer}, {Biazzo}, {Bignamini}, {Esposito}, {Giacobbe}, {H{\'e}brard}, {Malavolta}, {Maldonado}, {Mancini}, {Martinez Fiorenzano}, {Masiero}, {Nascimbeni}, {Pedani}, {Rainer}, \& {Scandariato}}]{Bonomo17}
{Bonomo}, A.~S., {Desidera}, S., {Benatti}, S., {et~al.} 2017, \aap, 602, A107, \dodoi{10.1051/0004-6361/201629882}

\bibitem[{{Chabrier} {et~al.}(2009){Chabrier}, {Baraffe}, {Leconte}, {Gallardo}, \& {Barman}}]{ChabrierBaraffe09}
{Chabrier}, G., {Baraffe}, I., {Leconte}, J., {Gallardo}, J., \& {Barman}, T. 2009, in American Institute of Physics Conference Series, Vol. 1094, 15th Cambridge Workshop on Cool Stars, Stellar Systems, and the Sun, ed. E.~{Stempels}, 102--111, \dodoi{10.1063/1.3099078}

\bibitem[{{Clausen} \& {Tilgner}(2015)}]{ClausenTilgner15}
{Clausen}, N., \& {Tilgner}, A. 2015, \aap, 584, A60, \dodoi{10.1051/0004-6361/201526082}

\bibitem[{{Correia} {et~al.}(2012){Correia}, {Bou{\'e}}, \& {Laskar}}]{correia12}
{Correia}, A. C.~M., {Bou{\'e}}, G., \& {Laskar}, J. 2012, \apjl, 744, L23, \dodoi{10.1088/2041-8205/744/2/L23}

\bibitem[{{Correia} {et~al.}(2020){Correia}, {Bourrier}, \& {Delisle}}]{correia20}
{Correia}, A.~C.~M., {Bourrier}, V., \& {Delisle}, J.~B. 2020, \aap, 635, A37, \dodoi{10.1051/0004-6361/201936967}

\bibitem[{{Cosentino} {et~al.}(2012){Cosentino}, {Lovis}, {Pepe}, {Collier Cameron}, {Latham}, {Molinari}, {Udry}, {Bezawada}, {Black}, {Born}, {Buchschacher}, {Charbonneau}, {Figueira}, {Fleury}, {Galli}, {Gallie}, {Gao}, {Ghedina}, {Gonzalez}, {Gonzalez}, {Guerra}, {Henry}, {Horne}, {Hughes}, {Kelly}, {Lodi}, {Lunney}, {Maire}, {Mayor}, {Micela}, {Ordway}, {Peacock}, {Phillips}, {Piotto}, {Pollacco}, {Queloz}, {Rice}, {Riverol}, {Riverol}, {San Juan}, {Sasselov}, {Segransan}, {Sozzetti}, {Sosnowska}, {Stobie}, {Szentgyorgyi}, {Vick}, \& {Weber}}]{harps-n12}
{Cosentino}, R., {Lovis}, C., {Pepe}, F., {et~al.} 2012, in Society of Photo-Optical Instrumentation Engineers (SPIE) Conference Series, Vol. 8446, Ground-based and Airborne Instrumentation for Astronomy IV, ed. I.~S. {McLean}, S.~K. {Ramsay}, \& H.~{Takami}, 84461V, \dodoi{10.1117/12.925738}

\bibitem[{{Cumming} {et~al.}(2008){Cumming}, {Butler}, {Marcy}, {Vogt}, {Wright}, \& {Fischer}}]{Cumming08}
{Cumming}, A., {Butler}, R.~P., {Marcy}, G.~W., {et~al.} 2008, \pasp, 120, 531, \dodoi{10.1086/588487}

\bibitem[{{Cumming} {et~al.}(1999){Cumming}, {Marcy}, \& {Butler}}]{Cumming99}
{Cumming}, A., {Marcy}, G.~W., \& {Butler}, R.~P. 1999, \apj, 526, 890, \dodoi{10.1086/308020}

\bibitem[{{Dawson} \& {Johnson}(2012)}]{dawson12}
{Dawson}, R.~I., \& {Johnson}, J.~A. 2012, \apj, 756, 122, \dodoi{10.1088/0004-637X/756/2/122}

\bibitem[{{Dawson} \& {Johnson}(2018)}]{dawson18}
---. 2018, \araa, 56, 175, \dodoi{10.1146/annurev-astro-081817-051853}

\bibitem[{{Deleuil} {et~al.}(2012){Deleuil}, {Bonomo}, {Ferraz-Mello}, {Erikson}, {Bouchy}, {Havel}, {Aigrain}, {Almenara}, {Alonso}, {Auvergne}, {Baglin}, {Barge}, {Bord{\'e}}, {Bruntt}, {Cabrera}, {Carpano}, {Cavarroc}, {Csizmadia}, {Damiani}, {Deeg}, {Dvorak}, {Fridlund}, {H{\'e}brard}, {Gandolfi}, {Gillon}, {Guenther}, {Guillot}, {Hatzes}, {Jorda}, {L{\'e}ger}, {Lammer}, {Mazeh}, {Moutou}, {Ollivier}, {Ofir}, {Parviainen}, {Queloz}, {Rauer}, {Rodr{\'\i}guez}, {Rouan}, {Santerne}, {Schneider}, {Tal-Or}, {Tingley}, {Weingrill}, \& {Wuchterl}}]{Deleuil12}
{Deleuil}, M., {Bonomo}, A.~S., {Ferraz-Mello}, S., {et~al.} 2012, \aap, 538, A145, \dodoi{10.1051/0004-6361/201117681}

\bibitem[{{Dong} {et~al.}(2021){Dong}, {Huang}, {Zhou}, {Dawson}, {Rodriguez}, {Eastman}, {Collins}, {Quinn}, {Shporer}, {Triaud}, {Wang}, {Beatty}, {Jackson}, {Collins}, {Abe}, {Suarez}, {Crouzet}, {M{\'e}karnia}, {Dransfield}, {Jensen}, {Stockdale}, {Barkaoui}, {Heitzmann}, {Wright}, {Addison}, {Wittenmyer}, {Okumura}, {Bowler}, {Horner}, {Kane}, {Kielkopf}, {Liu}, {Plavchan}, {Mengel}, {Ricker}, {Vanderspek}, {Latham}, {Seager}, {Winn}, {Jenkins}, {Christiansen}, \& {Paegert}}]{Dong21}
{Dong}, J., {Huang}, C.~X., {Zhou}, G., {et~al.} 2021, \apjl, 920, L16, \dodoi{10.3847/2041-8213/ac2600}

\bibitem[{{Fabrycky} \& {Tremaine}(2007)}]{fabrycky07}
{Fabrycky}, D., \& {Tremaine}, S. 2007, \apj, 669, 1298, \dodoi{10.1086/521702}

\bibitem[{{Feng} {et~al.}(2022){Feng}, {Butler}, {Vogt}, {Clement}, {Tinney}, {Cui}, {Aizawa}, {Jones}, {Bailey}, {Burt}, {Carter}, {Crane}, {Flammini Dotti}, {Holden}, {Ma}, {Ogihara}, {Oppenheimer}, {O'Toole}, {Shectman}, {Wittenmyer}, {Wang}, {Wright}, \& {Xuan}}]{feng22}
{Feng}, F., {Butler}, R.~P., {Vogt}, S.~S., {et~al.} 2022, \apjs, 262, 21, \dodoi{10.3847/1538-4365/ac7e57}

\bibitem[{{Ford} \& {Rasio}(2008)}]{Ford08}
{Ford}, E.~B., \& {Rasio}, F.~A. 2008, \apj, 686, 621, \dodoi{10.1086/590926}

\bibitem[{{Foreman-Mackey} {et~al.}(2013){Foreman-Mackey}, {Conley}, {Meierjurgen Farr}, {Hogg}, {Lang}, {Marshall}, {Price-Whelan}, {Sanders}, \& {Zuntz}}]{foreman-Mackey2013}
{Foreman-Mackey}, D., {Conley}, A., {Meierjurgen Farr}, W., {et~al.} 2013, {emcee: The MCMC Hammer}, Astrophysics Source Code Library.
\newblock \doeprint{1303.002}

\bibitem[{{Gaia Collaboration} {et~al.}(2021){Gaia Collaboration}, {Smart}, {Sarro}, {Rybizki}, {Reyl{\'e}}, {Robin}, {Hambly}, {Abbas}, {Barstow}, {de Bruijne}, {Bucciarelli}, {Carrasco}, {Cooper}, {Hodgkin}, {Masana}, {Michalik}, {Sahlmann}, {Sozzetti}, {Brown}, {Vallenari}, {Prusti}, {Babusiaux}, {Biermann}, {Creevey}, {Evans}, {Eyer}, {Hutton}, {Jansen}, {Jordi}, {Klioner}, {Lammers}, {Lindegren}, {Luri}, {Mignard}, {Panem}, {Pourbaix}, {Randich}, {Sartoretti}, {Soubiran}, {Walton}, {Arenou}, {Bailer-Jones}, {Bastian}, {Cropper}, {Drimmel}, {Katz}, {Lattanzi}, {van Leeuwen}, {Bakker}, {Casta{\~n}eda}, {De Angeli}, {Ducourant}, {Fabricius}, {Fouesneau}, {Fr{\'e}mat}, {Guerra}, {Guerrier}, {Guiraud}, {Jean-Antoine Piccolo}, {Messineo}, {Mowlavi}, {Nicolas}, {Nienartowicz}, {Pailler}, {Panuzzo}, {Riclet}, {Roux}, {Seabroke}, {Sordo}, {Tanga}, {Th{\'e}venin}, {Gracia-Abril}, {Portell}, {Teyssier}, {Altmann}, {Andrae}, {Bellas-Velidis}, {Benson}, {Berthier}, {Blomme}, {Brugaletta}, {Burgess}, {Busso}, {Carry},
  {Cellino}, {Cheek}, {Clementini}, {Damerdji}, {Davidson}, {Delchambre}, {Dell'Oro}, {Fern{\'a}ndez-Hern{\'a}ndez}, {Galluccio}, {Garc{\'\i}a-Lario}, {Garcia-Reinaldos}, {Gonz{\'a}lez-N{\'u}{\~n}ez}, {Gosset}, {Haigron}, {Halbwachs}, {Harrison}, {Hatzidimitriou}, {Heiter}, {Hern{\'a}ndez}, {Hestroffer}, {Holl}, {Jan{\ss}en}, {Jevardat de Fombelle}, {Jordan}, {Krone-Martins}, {Lanzafame}, {L{\"o}ffler}, {Lorca}, {Manteiga}, {Marchal}, {Marrese}, {Moitinho}, {Mora}, {Muinonen}, {Osborne}, {Pancino}, {Pauwels}, {Recio-Blanco}, {Richards}, {Riello}, {Rimoldini}, {Roegiers}, {Siopis}, {Smith}, {Ulla}, {Utrilla}, {van Leeuwen}, {van Reeven}, {Abreu Aramburu}, {Accart}, {Aerts}, {Aguado}, {Ajaj}, {Altavilla}, {{\'A}lvarez}, {{\'A}lvarez Cid-Fuentes}, {Alves}, {Anderson}, {Anglada Varela}, {Antoja}, {Audard}, {Baines}, {Baker}, {Balaguer-N{\'u}{\~n}ez}, {Balbinot}, {Balog}, {Barache}, {Barbato}, {Barros}, {Bartolom{\'e}}, {Bassilana}, {Bauchet}, {Baudesson-Stella}, {Becciani}, {Bellazzini}, {Bernet}, {Bertone},
  {Bianchi}, {Blanco-Cuaresma}, {Boch}, {Bombrun}, {Bossini}, {Bouquillon}, {Bragaglia}, {Bramante}, {Breedt}, {Bressan}, {Brouillet}, {Burlacu}, {Busonero}, {Butkevich}, {Buzzi}, {Caffau}, {Cancelliere}, {C{\'a}novas}, {Cantat-Gaudin}, {Carballo}, {Carlucci}, {Carnerero}, {Casamiquela}, {Castellani}, {Castro-Ginard}, {Castro Sampol}, {Chaoul}, {Charlot}, {Chemin}, {Chiavassa}, {Cioni}, {Comoretto}, {Cornez}, {Cowell}, {Crifo}, {Crosta}, {Crowley}, {Dafonte}, {Dapergolas}, {David}, {David}, {de Laverny}, {De Luise}, {De March}, {De Ridder}, {de Souza}, {de Teodoro}, {de Torres}, {del Peloso}, {del Pozo}, {Delgado}, {Delgado}, {Delisle}, {Di Matteo}, {Diakite}, {Diener}, {Distefano}, {Dolding}, {Eappachen}, {Edvardsson}, {Enke}, {Esquej}, {Fabre}, {Fabrizio}, {Faigler}, {Fedorets}, {Fernique}, {Fienga}, {Figueras}, {Fouron}, {Fragkoudi}, {Fraile}, {Franke}, {Gai}, {Garabato}, {Garcia-Gutierrez}, {Garc{\'\i}a-Torres}, {Garofalo}, {Gavras}, {Gerlach}, {Geyer}, {Giacobbe}, {Gilmore}, {Girona}, {Giuffrida},
  {Gomel}, {Gomez}, {Gonzalez-Santamaria}, {Gonz{\'a}lez-Vidal}, {Granvik}, {Guti{\'e}rrez-S{\'a}nchez}, {Guy}, {Hauser}, {Haywood}, {Helmi}, {Hidalgo}, {Hilger}, {H{\l}adczuk}, {Hobbs}, {Holland}, {Huckle}, {Jasniewicz}, {Jonker}, {Juaristi Campillo}, {Julbe}, {Karbevska}, {Kervella}, {Khanna}, {Kochoska}, {Kontizas}, {Kordopatis}, {Korn}, {Kostrzewa-Rutkowska}, {Kruszy{\'n}ska}, {Lambert}, {Lanza}, {Lasne}, {Le Campion}, {Le Fustec}, {Lebreton}, {Lebzelter}, {Leccia}, {Leclerc}, {Lecoeur-Taibi}, {Liao}, {Licata}, {Lindstr{\o}m}, {Lister}, {Livanou}, {Lobel}, {Madrero Pardo}, {Managau}, {Mann}, {Marchant}, {Marconi}, {Marcos Santos}, {Marinoni}, {Marocco}, {Marshall}, {Martin Polo}, {Mart{\'\i}n-Fleitas}, {Masip}, {Massari}, {Mastrobuono-Battisti}, {Mazeh}, {McMillan}, {Messina}, {Millar}, {Mints}, {Molina}, {Molinaro}, {Moln{\'a}r}, {Montegriffo}, {Mor}, {Morbidelli}, {Morel}, {Morris}, {Mulone}, {Munoz}, {Muraveva}, {Murphy}, {Musella}, {Noval}, {Ord{\'e}novic}, {Orr{\`u}}, {Osinde}, {Pagani}, {Pagano},
  {Palaversa}, {Palicio}, {Panahi}, {Pawlak}, {Pe{\~n}alosa Esteller}, {Penttil{\"a}}, {Piersimoni}, {Pineau}, {Plachy}, {Plum}, {Poggio}, {Poretti}, {Poujoulet}, {Pr{\v{s}}a}, {Pulone}, {Racero}, {Ragaini}, {Rainer}, {Raiteri}, {Rambaux}, {Ramos}, {Ramos-Lerate}, {Re Fiorentin}, {Regibo}, {Ripepi}, {Riva}, {Rixon}, {Robichon}, {Robin}, {Roelens}, {Rohrbasser}, {Romero-G{\'o}mez}, {Rowell}, {Royer}, {Rybicki}, {Sadowski}, {Sagrist{\`a} Sell{\'e}s}, {Salgado}, {Salguero}, {Samaras}, {Sanchez Gimenez}, {Sanna}, {Santove{\~n}a}, {Sarasso}, {Schultheis}, {Sciacca}, {Segol}, {Segovia}, {S{\'e}gransan}, {Semeux}, {Shahaf}, {Siddiqui}, {Siebert}, {Siltala}, {Slezak}, {Solano}, {Solitro}, {Souami}, {Souchay}, {Spagna}, {Spoto}, {Steele}, {Steidelm{\"u}ller}, {Stephenson}, {S{\"u}veges}, {Szabados}, {Szegedi-Elek}, {Taris}, {Tauran}, {Taylor}, {Teixeira}, {Thuillot}, {Tonello}, {Torra}, {Torra}, {Turon}, {Unger}, {Vaillant}, {van Dillen}, {Vanel}, {Vecchiato}, {Viala}, {Vicente}, {Voutsinas}, {Weiler}, {Wevers},
  {Wyrzykowski}, {Yoldas}, {Yvard}, {Zhao}, {Zorec}, {Zucker}, {Zurbach}, \& {Zwitter}}]{GCNS}
{Gaia Collaboration}, {Smart}, R.~L., {Sarro}, L.~M., {et~al.} 2021, \aap, 649, A6, \dodoi{10.1051/0004-6361/202039498}

\bibitem[{{Ghosh} \& {Chatterjee}(2023)}]{ghosh23}
{Ghosh}, T., \& {Chatterjee}, S. 2023, arXiv e-prints, arXiv:2304.12352, \dodoi{10.48550/arXiv.2304.12352}

\bibitem[{{Giacalone} {et~al.}(2017){Giacalone}, {Matsakos}, \& {K{\"o}nigl}}]{konigel17}
{Giacalone}, S., {Matsakos}, T., \& {K{\"o}nigl}, A. 2017, \aj, 154, 192, \dodoi{10.3847/1538-3881/aa8c04}

\bibitem[{{Goldberg} \& {Batygin}(2022)}]{goldberg22}
{Goldberg}, M., \& {Batygin}, K. 2022, \aj, 163, 201, \dodoi{10.3847/1538-3881/ac5961}

\bibitem[{{Goldreich} \& {Soter}(1966)}]{GoldreichSoter66}
{Goldreich}, P., \& {Soter}, S. 1966, \icarus, 5, 375, \dodoi{10.1016/0019-1035(66)90051-0}

\bibitem[{{Hara} {et~al.}(2019){Hara}, {Bou{\'e}}, {Laskar}, {Delisle}, \& {Unger}}]{hara19}
{Hara}, N.~C., {Bou{\'e}}, G., {Laskar}, J., {Delisle}, J.~B., \& {Unger}, N. 2019, \mnras, 489, 738, \dodoi{10.1093/mnras/stz1849}

\bibitem[{Harris {et~al.}(2020)Harris, Millman, van~der Walt, Gommers, Virtanen, Cournapeau, Wieser, Taylor, Berg, Smith, Kern, Picus, Hoyer, van Kerkwijk, Brett, Haldane, del R{\'{i}}o, Wiebe, Peterson, G{\'{e}}rard-Marchant, Sheppard, Reddy, Weckesser, Abbasi, Gohlke, \& Oliphant}]{NumPy}
Harris, C.~R., Millman, K.~J., van~der Walt, S.~J., {et~al.} 2020, Nature, 585, 357, \dodoi{10.1038/s41586-020-2649-2}

\bibitem[{{He} {et~al.}(2020){He}, {Ford}, {Ragozzine}, \& {Carrera}}]{he20}
{He}, M.~Y., {Ford}, E.~B., {Ragozzine}, D., \& {Carrera}, D. 2020, \aj, 160, 276, \dodoi{10.3847/1538-3881/abba18}

\bibitem[{{Hebb} {et~al.}(2009){Hebb}, {Collier-Cameron}, {Loeillet}, {Pollacco}, {H{\'e}brard}, {Street}, {Bouchy}, {Stempels}, {Moutou}, {Simpson}, {Udry}, {Joshi}, {West}, {Skillen}, {Wilson}, {McDonald}, {Gibson}, {Aigrain}, {Anderson}, {Benn}, {Christian}, {Enoch}, {Haswell}, {Hellier}, {Horne}, {Irwin}, {Lister}, {Maxted}, {Mayor}, {Norton}, {Parley}, {Pont}, {Queloz}, {Smalley}, \& {Wheatley}}]{Hebb09}
{Hebb}, L., {Collier-Cameron}, A., {Loeillet}, B., {et~al.} 2009, \apj, 693, 1920, \dodoi{10.1088/0004-637X/693/2/1920}

\bibitem[{Hunter(2007)}]{Matplotlib}
Hunter, J.~D. 2007, Computing in Science \& Engineering, 9, 90, \dodoi{10.1109/MCSE.2007.55}

\bibitem[{{Husnoo} {et~al.}(2012){Husnoo}, {Pont}, {Mazeh}, {Fabrycky}, {H{\'e}brard}, {Bouchy}, \& {Shporer}}]{pont12}
{Husnoo}, N., {Pont}, F., {Mazeh}, T., {et~al.} 2012, \mnras, 422, 3151, \dodoi{10.1111/j.1365-2966.2012.20839.x}

\bibitem[{{Ida} \& {Lin}(2004)}]{ida04}
{Ida}, S., \& {Lin}, D.~N.~C. 2004, \apj, 604, 388, \dodoi{10.1086/381724}

\bibitem[{{Jackson} {et~al.}(2023){Jackson}, {Dawson}, {Quarles}, \& {Dong}}]{Jackson23}
{Jackson}, J.~M., {Dawson}, R.~I., {Quarles}, B., \& {Dong}, J. 2023, \aj, 165, 82, \dodoi{10.3847/1538-3881/acac86}

\bibitem[{{Kane} {et~al.}(2012){Kane}, {Ciardi}, {Gelino}, \& {von Braun}}]{kane12}
{Kane}, S.~R., {Ciardi}, D.~R., {Gelino}, D.~M., \& {von Braun}, K. 2012, \mnras, 425, 757, \dodoi{10.1111/j.1365-2966.2012.21627.x}

\bibitem[{{Ke-ting} {et~al.}(2024){Ke-ting}, {Dong-sheng}, {Ji-wei}, \& {Ji-lin}}]{ke-ting24}
{Ke-ting}, S., {Dong-sheng}, A., {Ji-wei}, X., \& {Ji-lin}, Z. 2024, \caa, 48, 1, \dodoi{10.1016/j.chinastron.2024.03.006}

\bibitem[{{Kipping}(2008)}]{kipping08}
{Kipping}, D.~M. 2008, \mnras, 389, 1383, \dodoi{10.1111/j.1365-2966.2008.13658.x}

\bibitem[{{Kipping}(2013)}]{kipping13}
---. 2013, \mnras, 434, L51, \dodoi{10.1093/mnrasl/slt075}

\bibitem[{{Kipping}(2014)}]{kipping14}
---. 2014, \mnras, 444, 2263, \dodoi{10.1093/mnras/stu1561}

\bibitem[{{Lissauer} {et~al.}(2011){Lissauer}, {Ragozzine}, {Fabrycky}, {Steffen}, {Ford}, {Jenkins}, {Shporer}, {Holman}, {Rowe}, {Quintana}, {Batalha}, {Borucki}, {Bryson}, {Caldwell}, {Carter}, {Ciardi}, {Dunham}, {Fortney}, {Gautier}, {Howell}, {Koch}, {Latham}, {Marcy}, {Morehead}, \& {Sasselov}}]{lissauer11}
{Lissauer}, J.~J., {Ragozzine}, D., {Fabrycky}, D.~C., {et~al.} 2011, \apjs, 197, 8, \dodoi{10.1088/0067-0049/197/1/8}

\bibitem[{{Lucy} \& {Sweeney}(1971)}]{LucySweeney71}
{Lucy}, L.~B., \& {Sweeney}, M.~A. 1971, \aj, 76, 544, \dodoi{10.1086/111159}

\bibitem[{{Mardling} \& {Lin}(2002)}]{MardlingLin02}
{Mardling}, R.~A., \& {Lin}, D.~N.~C. 2002, \apj, 573, 829, \dodoi{10.1086/340752}

\bibitem[{{Matsumoto} {et~al.}(2015){Matsumoto}, {Nagasawa}, \& {Ida}}]{matsumoto15}
{Matsumoto}, Y., {Nagasawa}, M., \& {Ida}, S. 2015, \apj, 810, 106, \dodoi{10.1088/0004-637X/810/2/106}

\bibitem[{{Matsumura} {et~al.}(2010){Matsumura}, {Peale}, \& {Rasio}}]{MatsumuraRasio10}
{Matsumura}, S., {Peale}, S.~J., \& {Rasio}, F.~A. 2010, \apj, 725, 1995, \dodoi{10.1088/0004-637X/725/2/1995}

\bibitem[{{Matsumura} {et~al.}(2008){Matsumura}, {Takeda}, \& {Rasio}}]{MatsumaraTakedaRasio09}
{Matsumura}, S., {Takeda}, G., \& {Rasio}, F.~A. 2008, \apjl, 686, L29, \dodoi{10.1086/592818}

\bibitem[{{Mazeh} {et~al.}(2016){Mazeh}, {Holczer}, \& {Faigler}}]{mazeh16}
{Mazeh}, T., {Holczer}, T., \& {Faigler}, S. 2016, \aap, 589, A75, \dodoi{10.1051/0004-6361/201528065}

\bibitem[{{Mazeh} \& {Shaham}(1979)}]{mazeh79}
{Mazeh}, T., \& {Shaham}, J. 1979, \aap, 77, 145

\bibitem[{{Mills} {et~al.}(2019){Mills}, {Howard}, {Petigura}, {Fulton}, {Isaacson}, \& {Weiss}}]{Mills19}
{Mills}, S.~M., {Howard}, A.~W., {Petigura}, E.~A., {et~al.} 2019, \aj, 157, 198, \dodoi{10.3847/1538-3881/ab1009}

\bibitem[{{Mishra} {et~al.}(2023){Mishra}, {Alibert}, {Udry}, \& {Mordasini}}]{udry23}
{Mishra}, L., {Alibert}, Y., {Udry}, S., \& {Mordasini}, C. 2023, \aap, 670, A68, \dodoi{10.1051/0004-6361/202243751}

\bibitem[{{Mordasini} \& {Burn}(2024)}]{mordasini24}
{Mordasini}, C., \& {Burn}, R. 2024, arXiv e-prints, arXiv:2404.15555, \dodoi{10.48550/arXiv.2404.15555}

\bibitem[{{Murray} \& {Dermott}(1999)}]{MurrayDermott99}
{Murray}, C.~D., \& {Dermott}, S.~F. 1999, {Solar system dynamics}

\bibitem[{{Naef} {et~al.}(2001){Naef}, {Latham}, {Mayor}, {Mazeh}, {Beuzit}, {Drukier}, {Perrier-Bellet}, {Queloz}, {Sivan}, {Torres}, {Udry}, \& {Zucker}}]{Naef01}
{Naef}, D., {Latham}, D.~W., {Mayor}, M., {et~al.} 2001, \aap, 375, L27, \dodoi{10.1051/0004-6361:20010853}

\bibitem[{{Obertas} {et~al.}(2023){Obertas}, {Tamayo}, \& {Murray}}]{murray23}
{Obertas}, A., {Tamayo}, D., \& {Murray}, N. 2023, \mnras, \dodoi{10.1093/mnras/stad1921}

\bibitem[{{O'Toole} {et~al.}(2009){O'Toole}, {Tinney}, {Jones}, {Butler}, {Marcy}, {Carter}, \& {Bailey}}]{otoole09}
{O'Toole}, S.~J., {Tinney}, C.~G., {Jones}, H.~R.~A., {et~al.} 2009, \mnras, 392, 641, \dodoi{10.1111/j.1365-2966.2008.14051.x}

\bibitem[{{Papaloizou}(2011)}]{papaloizou11}
{Papaloizou}, J.~C.~B. 2011, Celestial Mechanics and Dynamical Astronomy, 111, 83, \dodoi{10.1007/s10569-011-9344-4}

\bibitem[{Pedregosa {et~al.}(2011)Pedregosa, Varoquaux, Gramfort, Michel, Thirion, Grisel, Blondel, Prettenhofer, Weiss, Dubourg, {et~al.}}]{pedregosa2011scikit}
Pedregosa, F., Varoquaux, G., Gramfort, A., {et~al.} 2011, Journal of machine learning research, 12, 2825

\bibitem[{{Pepe} {et~al.}(2021){Pepe}, {Cristiani}, {Rebolo}, {Santos}, {Dekker}, {Cabral}, {Di Marcantonio}, {Figueira}, {Lo Curto}, {Lovis}, {Mayor}, {M{\'e}gevand}, {Molaro}, {Riva}, {Zapatero Osorio}, {Amate}, {Manescau}, {Pasquini}, {Zerbi}, {Adibekyan}, {Abreu}, {Affolter}, {Alibert}, {Aliverti}, {Allart}, {Allende Prieto}, {{\'A}lvarez}, {Alves}, {Avila}, {Baldini}, {Bandy}, {Barros}, {Benz}, {Bianco}, {Borsa}, {Bourrier}, {Bouchy}, {Broeg}, {Calderone}, {Cirami}, {Coelho}, {Conconi}, {Coretti}, {Cumani}, {Cupani}, {D'Odorico}, {Damasso}, {Deiries}, {Delabre}, {Demangeon}, {Dumusque}, {Ehrenreich}, {Faria}, {Fragoso}, {Genolet}, {Genoni}, {G{\'e}nova Santos}, {Gonz{\'a}lez Hern{\'a}ndez}, {Hughes}, {Iwert}, {Kerber}, {Knudstrup}, {Landoni}, {Lavie}, {Lillo-Box}, {Lizon}, {Maire}, {Martins}, {Mehner}, {Micela}, {Modigliani}, {Monteiro}, {Monteiro}, {Moschetti}, {Murphy}, {Nunes}, {Oggioni}, {Oliveira}, {Oshagh}, {Pall{\'e}}, {Pariani}, {Poretti}, {Rasilla}, {Rebord{\~a}o}, {Redaelli}, {Santana Tschudi},
  {Santin}, {Santos}, {S{\'e}gransan}, {Schmidt}, {Segovia}, {Sosnowska}, {Sozzetti}, {Sousa}, {Span{\`o}}, {Su{\'a}rez Mascare{\~n}o}, {Tabernero}, {Tenegi}, {Udry}, \& {Zanutta}}]{espresso21}
{Pepe}, F., {Cristiani}, S., {Rebolo}, R., {et~al.} 2021, \aap, 645, A96, \dodoi{10.1051/0004-6361/202038306}

\bibitem[{{Perryman}(2018)}]{Perryman18}
{Perryman}, M. 2018, {The Exoplanet Handbook}

\bibitem[{{Petrovich} {et~al.}(2014){Petrovich}, {Tremaine}, \& {Rafikov}}]{petrovich14}
{Petrovich}, C., {Tremaine}, S., \& {Rafikov}, R. 2014, \apj, 786, 101, \dodoi{10.1088/0004-637X/786/2/101}

\bibitem[{{Pollacco} {et~al.}(2006){Pollacco}, {Skillen}, {Collier Cameron}, {Christian}, {Hellier}, {Irwin}, {Lister}, {Street}, {West}, {Anderson}, {Clarkson}, {Deeg}, {Enoch}, {Evans}, {Fitzsimmons}, {Haswell}, {Hodgkin}, {Horne}, {Kane}, {Keenan}, {Maxted}, {Norton}, {Osborne}, {Parley}, {Ryans}, {Smalley}, {Wheatley}, \& {Wilson}}]{WASP06}
{Pollacco}, D.~L., {Skillen}, I., {Collier Cameron}, A., {et~al.} 2006, \pasp, 118, 1407, \dodoi{10.1086/508556}

\bibitem[{{Pont} {et~al.}(2011){Pont}, {Husnoo}, {Mazeh}, \& {Fabrycky}}]{pont11}
{Pont}, F., {Husnoo}, N., {Mazeh}, T., \& {Fabrycky}, D. 2011, \mnras, 414, 1278, \dodoi{10.1111/j.1365-2966.2011.18462.x}

\bibitem[{{Quirrenbach} {et~al.}(2014){Quirrenbach}, {Amado}, {Caballero}, {Mundt}, {Reiners}, {Ribas}, {Seifert}, {Abril}, {Aceituno}, {Alonso-Floriano}, {Ammler-von Eiff}, {Antona Jim{\'e}nez}, {Anwand-Heerwart}, {Azzaro}, {Bauer}, {Barrado}, {Becerril}, {B{\'e}jar}, {Ben{\'\i}tez}, {Berdi{\~n}as}, {C{\'a}rdenas}, {Casal}, {Claret}, {Colom{\'e}}, {Cort{\'e}s-Contreras}, {Czesla}, {Doellinger}, {Dreizler}, {Feiz}, {Fern{\'a}ndez}, {Galad{\'\i}}, {G{\'a}lvez-Ortiz}, {Garc{\'\i}a-Piquer}, {Garc{\'\i}a-Vargas}, {Garrido}, {Gesa}, {G{\'o}mez Galera}, {Gonz{\'a}lez {\'A}lvarez}, {Gonz{\'a}lez Hern{\'a}ndez}, {Gr{\"o}zinger}, {Gu{\`a}rdia}, {Guenther}, {de Guindos}, {Guti{\'e}rrez-Soto}, {Hagen}, {Hatzes}, {Hauschildt}, {Helmling}, {Henning}, {Hermann}, {Hern{\'a}ndez Casta{\~n}o}, {Herrero}, {Hidalgo}, {Holgado}, {Huber}, {Huber}, {Jeffers}, {Joergens}, {de Juan}, {Kehr}, {Klein}, {K{\"u}rster}, {Lamert}, {Lalitha}, {Laun}, {Lemke}, {Lenzen}, {L{\'o}pez del Fresno}, {L{\'o}pez Mart{\'\i}}, {L{\'o}pez-Santiago},
  {Mall}, {Mandel}, {Mart{\'\i}n}, {Mart{\'\i}n-Ruiz}, {Mart{\'\i}nez-Rodr{\'\i}guez}, {Marvin}, {Mathar}, {Mirabet}, {Montes}, {Morales Mu{\~n}oz}, {Moya}, {Naranjo}, {Ofir}, {Oreiro}, {Pall{\'e}}, {Panduro}, {Passegger}, {P{\'e}rez-Calpena}, {P{\'e}rez Medialdea}, {Perger}, {Pluto}, {Ram{\'o}n}, {Rebolo}, {Redondo}, {Reffert}, {Reinhardt}, {Rhode}, {Rix}, {Rodler}, {Rodr{\'\i}guez}, {Rodr{\'\i}guez-L{\'o}pez}, {Rodr{\'\i}guez-P{\'e}rez}, {Rohloff}, {Rosich}, {S{\'a}nchez-Blanco}, {S{\'a}nchez Carrasco}, {Sanz-Forcada}, {Sarmiento}, {Sch{\"a}fer}, {Schiller}, {Schmidt}, {Schmitt}, {Solano}, {Stahl}, {Storz}, {St{\"u}rmer}, {Su{\'a}rez}, {Ulbrich}, {Veredas}, {Wagner}, {Winkler}, {Zapatero Osorio}, {Zechmeister}, {Abell{\'a}n de Paco}, {Anglada-Escud{\'e}}, {del Burgo}, {Klutsch}, {Lizon}, {L{\'o}pez-Morales}, {Morales}, {Perryman}, {Tulloch}, \& {Xu}}]{carmenes14}
{Quirrenbach}, A., {Amado}, P.~J., {Caballero}, J.~A., {et~al.} 2014, in Society of Photo-Optical Instrumentation Engineers (SPIE) Conference Series, Vol. 9147, Ground-based and Airborne Instrumentation for Astronomy V, ed. S.~K. {Ramsay}, I.~S. {McLean}, \& H.~{Takami}, 91471F, \dodoi{10.1117/12.2056453}

\bibitem[{{Rasio} {et~al.}(1996){Rasio}, {Tout}, {Lubow}, \& {Livio}}]{rasio96}
{Rasio}, F.~A., {Tout}, C.~A., {Lubow}, S.~H., \& {Livio}, M. 1996, \apj, 470, 1187, \dodoi{10.1086/177941}

\bibitem[{{Rey} {et~al.}(2018){Rey}, {Bouchy}, {Stalport}, {Deleuil}, {H{\'e}brard}, {Almenara}, {Alonso}, {Barros}, {Bonomo}, {Cazalet}, {Delisle}, {D{\'\i}az}, {Fridlund}, {Guenther}, {Guillot}, {Montagnier}, {Moutou}, {Lovis}, {Queloz}, {Santerne}, \& {Udry}}]{Rey18}
{Rey}, J., {Bouchy}, F., {Stalport}, M., {et~al.} 2018, \aap, 619, A115, \dodoi{10.1051/0004-6361/201833180}

\bibitem[{{Ribas} \& {Miralda-Escud{\'e}}(2007)}]{Ribas&Miralda07}
{Ribas}, I., \& {Miralda-Escud{\'e}}, J. 2007, \aap, 464, 779, \dodoi{10.1051/0004-6361:20065726}

\bibitem[{{Rosenthal} {et~al.}(2024){Rosenthal}, {Howard}, {Knutson}, \& {Fulton}}]{Rosenthal24}
{Rosenthal}, L.~J., {Howard}, A.~W., {Knutson}, H.~A., \& {Fulton}, B.~J. 2024, \apjs, 270, 1, \dodoi{10.3847/1538-4365/acffc0}

\bibitem[{{Sagear} \& {Ballard}(2023)}]{ballard23}
{Sagear}, S., \& {Ballard}, S. 2023, Proceedings of the National Academy of Science, 120, e2217398120, \dodoi{10.1073/pnas.2217398120}

\bibitem[{{Shabram} {et~al.}(2016){Shabram}, {Demory}, {Cisewski}, {Ford}, \& {Rogers}}]{shabram16}
{Shabram}, M., {Demory}, B.-O., {Cisewski}, J., {Ford}, E.~B., \& {Rogers}, L. 2016, \apj, 820, 93, \dodoi{10.3847/0004-637X/820/2/93}

\bibitem[{{Socrates} {et~al.}(2012){Socrates}, {Katz}, {Dong}, \& {Tremaine}}]{Socrates12}
{Socrates}, A., {Katz}, B., {Dong}, S., \& {Tremaine}, S. 2012, \apj, 750, 106, \dodoi{10.1088/0004-637X/750/2/106}

\bibitem[{{Sozzetti} {et~al.}(2023){Sozzetti}, {Pinamonti}, {Damasso}, {Desidera}, {Biazzo}, {Bonomo}, {Nardiello}, {Gratton}, {Lanza}, {Malavolta}, {Giacobbe}, {Affer}, {Bignamini}, {Borsa}, {Boschin}, {Brogi}, {Cabona}, {Claudi}, {Covino}, {Di Fabrizio}, {Ghedina}, {Harutyunyan}, {Knapic}, {Maldonado}, {Maggio}, {Mancini}, {Mantovan}, {Marzari}, {Messina}, {Micela}, {Molinari}, {Montalto}, {Naponiello}, {Pagano}, {Pedani}, {Piotto}, {Poretti}, {Scandariato}, {Silvotti}, \& {Turrini}}]{sozzetti23}
{Sozzetti}, A., {Pinamonti}, M., {Damasso}, M., {et~al.} 2023, \aap, 677, L15, \dodoi{10.1051/0004-6361/202347329}

\bibitem[{{Tang} {et~al.}(2024){Tang}, {Li}, {Xiao}, {Gai}, {Li}, {Dong}, {Wang}, \& {Gao}}]{Tang24}
{Tang}, Y., {Li}, X., {Xiao}, K., {et~al.} 2024, Universe, 10, 182, \dodoi{10.3390/universe10040182}

\bibitem[{{Tremaine}(2015)}]{tremaine15}
{Tremaine}, S. 2015, \apj, 807, 157, \dodoi{10.1088/0004-637X/807/2/157}

\bibitem[{{Udry} {et~al.}(2003){Udry}, {Mayor}, \& {Santos}}]{Udry03}
{Udry}, S., {Mayor}, M., \& {Santos}, N.~C. 2003, \aap, 407, 369, \dodoi{10.1051/0004-6361:20030843}

\bibitem[{{Ulmer-Moll} {et~al.}(2019){Ulmer-Moll}, {Santos}, {Figueira}, {Brinchmann}, \& {Faria}}]{Ulmer-Moll19}
{Ulmer-Moll}, S., {Santos}, N.~C., {Figueira}, P., {Brinchmann}, J., \& {Faria}, J.~P. 2019, \aap, 630, A135, \dodoi{10.1051/0004-6361/201936049}

\bibitem[{{Van Eylen} \& {Albrecht}(2015)}]{EylenAlbrecht15}
{Van Eylen}, V., \& {Albrecht}, S. 2015, \apj, 808, 126, \dodoi{10.1088/0004-637X/808/2/126}

\bibitem[{{Van Eylen} {et~al.}(2016){Van Eylen}, {Winn}, \& {Albrecht}}]{EylenWinnAlbrecht16}
{Van Eylen}, V., {Winn}, J.~N., \& {Albrecht}, S. 2016, \apj, 824, 15, \dodoi{10.3847/0004-637X/824/1/15}

\bibitem[{{Van Eylen} {et~al.}(2019){Van Eylen}, {Albrecht}, {Huang}, {MacDonald}, {Dawson}, {Cai}, {Foreman-Mackey}, {Lundkvist}, {Silva Aguirre}, {Snellen}, \& {Winn}}]{Eylen19}
{Van Eylen}, V., {Albrecht}, S., {Huang}, X., {et~al.} 2019, \aj, 157, 61, \dodoi{10.3847/1538-3881/aaf22f}

\bibitem[{{Vick} {et~al.}(2023){Vick}, {Su}, \& {Lai}}]{vick23}
{Vick}, M., {Su}, Y., \& {Lai}, D. 2023, \apjl, 943, L13, \dodoi{10.3847/2041-8213/acaea6}

\bibitem[{Virtanen {et~al.}(2020)Virtanen, Gommers, Oliphant, Haberland, Reddy, Cournapeau, Burovski, Peterson, Weckesser, Bright, {van der Walt}, Brett, Wilson, Millman, Mayorov, Nelson, Jones, Kern, Larson, Carey, Polat, Feng, Moore, {VanderPlas}, Laxalde, Perktold, Cimrman, Henriksen, Quintero, Harris, Archibald, Ribeiro, Pedregosa, {van Mulbregt}, \& {SciPy 1.0 Contributors}}]{SciPy}
Virtanen, P., Gommers, R., Oliphant, T.~E., {et~al.} 2020, Nature Methods, 17, 261, \dodoi{10.1038/s41592-019-0686-2}

\bibitem[{{Wang} {et~al.}(2019){Wang}, {Wang}, {Hinse}, {Wu}, {Davis}, {Hori}, {Yoon}, {Han}, {Nie}, {Liu}, {Zhang}, {Zhou}, {Wittenmyer}, {Peng}, \& {Laughlin}}]{Wang19}
{Wang}, Y.-H., {Wang}, S., {Hinse}, T.~C., {et~al.} 2019, \aj, 157, 82, \dodoi{10.3847/1538-3881/aaf6b6}

\bibitem[{{Weiss} {et~al.}(2023){Weiss}, {Millholland}, {Petigura}, {Adams}, {Batygin}, {Block}, \& {Mordasini}}]{weiss23}
{Weiss}, L.~M., {Millholland}, S.~C., {Petigura}, E.~A., {et~al.} 2023, in Astronomical Society of the Pacific Conference Series, Vol. 534, Astronomical Society of the Pacific Conference Series, ed. S.~{Inutsuka}, Y.~{Aikawa}, T.~{Muto}, K.~{Tomida}, \& M.~{Tamura}, 863

\bibitem[{{Wittenmyer} {et~al.}(2019){Wittenmyer}, {Bergmann}, {Horner}, {Clark}, \& {Kane}}]{Wittenmyer19}
{Wittenmyer}, R.~A., {Bergmann}, C., {Horner}, J., {Clark}, J., \& {Kane}, S.~R. 2019, \mnras, 484, 4230, \dodoi{10.1093/mnras/stz236}

\bibitem[{{Xie} {et~al.}(2016){Xie}, {Dong}, {Zhu}, {Huber}, {Zheng}, {De Cat}, {Fu}, {Liu}, {Luo}, {Wu}, {Zhang}, {Zhang}, {Zhou}, {Cao}, {Hou}, {Wang}, \& {Zhang}}]{XieDong16}
{Xie}, J.-W., {Dong}, S., {Zhu}, Z., {et~al.} 2016, Proceedings of the National Academy of Science, 113, 11431, \dodoi{10.1073/pnas.1604692113}

\bibitem[{{Yang} \& {Jin}(2024)}]{Yang24}
{Yang}, J., \& {Jin}, L. 2024, \aap, 685, A35, \dodoi{10.1051/0004-6361/202348347}

\bibitem[{{Yee} {et~al.}(2020){Yee}, {Winn}, {Knutson}, {Patra}, {Vissapragada}, {Zhang}, {Holman}, {Shporer}, \& {Wright}}]{YeeWinn20}
{Yee}, S.~W., {Winn}, J.~N., {Knutson}, H.~A., {et~al.} 2020, \apjl, 888, L5, \dodoi{10.3847/2041-8213/ab5c16}

\bibitem[{{Zhu} \& {Dong}(2021)}]{ZhuDong21ARAA}
{Zhu}, W., \& {Dong}, S. 2021, \araa, 59, 291, \dodoi{10.1146/annurev-astro-112420-020055}

\end{thebibliography}
\bibliographystyle{aasjournal}



\end{document}